\documentclass[preprint]{aastex}

\usepackage{epsfig}

\shorttitle{Marconi et al.}
\shortauthors{Black Hole at the Nucleus of Centaurus A}

\newcommand{\Xo}{\ensuremath{x_0}}
\newcommand{\Xok}{\ensuremath{x_{0k}}}
\newcommand{\B}{\ensuremath{b}}
\newcommand{\Th}{\ensuremath{\theta}}
\newcommand{\I}{\ensuremath{i}}
\newcommand{\M}{\ensuremath{M_\bullet}}
\newcommand{\Vs}{\ensuremath{V_\mathrm{Sys}}}

\newcommand{\Chi}{\ensuremath{\chi_\mathrm{Red}^2}}
\newcommand{\xten}[1]{\ensuremath{\times 10^{#1}}}
\newcommand{\ten}[1]{\ensuremath{10^{#1}}}
\newcommand{\HMOL}{\ensuremath{\mathrm{H}_2}}
\newcommand{\HA}{\ensuremath{\mathrm{H}\alpha}}
\newcommand{\PA}{\ensuremath{\mathrm{Pa}\alpha}}
\newcommand{\PB}{\ensuremath{\mathrm{Pa}\beta}}

\newcommand{\BG}{\ensuremath{\mathrm{Br}\gamma}}

\newcommand{\AV}{\ensuremath{A_\mathrm{V}}}

\newcommand{\Lo}{\ensuremath{\mathrm{L}_\odot}}
\newcommand{\LVo}{\ensuremath{\mathrm{L}_{\mathrm{V}\odot}}}
\newcommand{\LKo}{\ensuremath{\mathrm{L}_{\mathrm{K}\odot}}}

\newcommand{\Mo}{\ensuremath{\mathrm{M}_\odot}}
\newcommand{\LAGN}{\ensuremath{L_\mathrm{AGN}}}

\newcommand{\LV}{\ensuremath{L_\mathrm{V}}}
\newcommand{\MV}{\ensuremath{M_\mathrm{V}}}

\newcommand{\LFIR}{\ensuremath{L_\mathrm{FIR}}}

\newcommand{\Ne}{\ensuremath{N_\mathrm{e}}}
\newcommand{\Te}{\ensuremath{T_\mathrm{e}}}
\newcommand{\EV}{\ensuremath{\,\mathrm{eV}}}
\newcommand{\KEV}{\ensuremath{\,\mathrm{keV}}}
\newcommand{\ERG}{\ensuremath{\,\mathrm{erg}}}

\renewcommand{\S}{\ensuremath{\,\mathrm{s}}}

\newcommand{\MPC}{\ensuremath{\,\mathrm{Mpc}}}
\newcommand{\KPC}{\ensuremath{\,\mathrm{kpc}}}
\newcommand{\PC}{\ensuremath{\,\mathrm{pc}}}
\newcommand{\KM}{\ensuremath{\,\mathrm{km}}}
\newcommand{\CM}{\ensuremath{\,\mathrm{cm}}}
\newcommand{\MIC}{\ensuremath{\,\mathrm{\mu m}}}

\newcommand{\1}{\ensuremath{^{-1}}}

\newcommand{\3}{\ensuremath{^{-3}}}
\newcommand{\forb}[2]{\ensuremath{[\mathrm{#1}\,\textsc{\lowercase{#2}}]}}
\newcommand{\perm}[2]{\ensuremath{\mathrm{#1}\,\textsc{\lowercase{#2}}}}
\newcommand{\WL}{\ensuremath{\lambda}}

\newcommand{\SIX}{\forb{S}{IX}}
\newcommand{\FeII}{\forb{Fe}{II}}
\newcommand{\NII}{\forb{N}{II}}
\newcommand{\HeI}{\perm{He}{I}}

\begin{document}

\title{Peering through the dust: Evidence for a
supermassive Black Hole at the nucleus of Centaurus A from VLT IR 
spectroscopy \altaffilmark{1}}

\author{Alessandro Marconi}
\affil{Osservatorio Astrofisico di Arcetri\\
       Largo E. Fermi 5,\\
       I-50125 Firenze, ITALY}
\email{marconi@arcetri.astro.it}

\author{Alessandro Capetti}
\affil{Osservatorio Astronomico di Torino\\
       Strada Osservatorio 20,\\
       I-10025 Pino Torinese, ITALY}
\email{capetti@to.astro.it }

\author{David Axon}
\affil{Department of Physical Sciences\\
       University of Hertfordshire\\
       College Lane,\\
       Hatfield, Herts AL10 9AB, UK}
\email{dja@star.herts.ac.uk}

\author{Anton Koekemoer, Duccio Macchetto\altaffilmark{2}, Ethan J. 
Schreier}
\affil{Space Telescope Science Institute\\
       3700 San Martin Drive,\\
       Baltimore, MD 21218, USA}
\email{koekemoe@stsci.edu, macchetto@stsci.edu, schreier@stsci.edu}

\altaffiltext{1}{Based on observations collected at the European Southern
Observatory, Paranal, Chile (ESO Program 63.P-0271A)}
\altaffiltext{2}{Affiliated to Space Science Department of ESA}

\clearpage
\begin{abstract}
We used the near infrared spectrometer ISAAC at the ESO
{\it Very Large Telescope} to map the velocity field of
Centaurus A (NGC 5128) at several position angles and locations
in the central 20\arcsec\ of the galaxy. The high spatial resolution ($\sim 
0\farcs5$) velocity fields
from both ionized and molecular gas (\PB, \FeII, \BG\ and \HMOL)
are not compromised by either excitation effects or obscuration.
We identify three distinct kinematical systems:
(i) a rotating "nuclear disk" of ionized gas, confined
to the inner 2\arcsec, the counterpart of the
\PA\ feature previously revealed by HST/NICMOS imaging;
(ii) a ring-like system with a $\sim 6\arcsec$ inner radius
detected only in \HMOL\, likely the counterpart
of the 100\PC-scale structure detected in CO by other authors;
(iii) a normal extended component of gas rotating in the galactic potential.
The nuclear disk is in keplerian rotation around a central mass
concentration, dark ($M/L> 20\,\Mo/\LKo$) and point-like at the spatial
resolution of the data ($R<0\farcs25\sim 4\PC$). We interpret this mass 
concentration
as a supermassive black hole. Its dynamical mass based on the 
line velocities and disk inclination ($\I>15\arcdeg$)
is $\M = 2^{+3.0}_{-1.4}\xten{8}\Mo$.
The ring-like system is probably characterized by non-circular motions; a 
"figure-of-8" pattern observed in the \HMOL\ position-velocity diagram
might provide kinematical evidence for the presence of a nuclear bar.
\end{abstract}


\vskip 0.5cm
\section{Introduction}

Mass accretion onto a supermassive black hole (BH) is widely
accepted as the most likely mechanism for powering Active
Galactic Nuclei (AGN).  It provides a simple and efficient way
to produce the necessary large amount of energy in a small volume. 
Demonstrating the existence of supermassive black holes and
measuring their masses has thus been a crucial element in confirming
the standard view of AGNs.

Over the last few years, there have been direct dynamical mass 
determinations for BHs in nearby galaxies in sufficient numbers
and with sufficient accuracy to enable detailed demographic studies 
\citep[e.g. the recent reviews by][]{macchetto99,ho98,ford97,marel97,kr95}.
The possible correlation observed between
the mass of the BH and that of the host spheroid (either the bulge of a 
spiral galaxy, or the entire mass of an elliptical galaxy)
has prompted serious studies of the relationship between BH growth
and host galaxy evolution \citep{magorrian,rich98}.

The nuclei of many AGNs and normal galaxies alike are heavily obscured by dust,
making it difficult to study kinematics at visible wavelengths. For example,
the 3CR HST snapshot survey has shown that the nuclei of radio galaxies are
often covered by foreground dust lanes \citep{dekoff96,martel99}.  This has
been further confirmed by other sensitive surveys with WFPC2 and NICMOS
\citep{malkan,regan}.  Such dust obscuration can distort both absorption
line and emission line velocity field measurements to such an extent that
reliable BH mass estimates are difficult or impossible using optical data
alone.  IR spectroscopy is thus increasingly important to continue to find and
measure the masses of black holes in active galaxies.

Centaurus A (NGC5128), the closest active giant elliptical galaxy,
nicely illustrates the problem of dust obscuration. At optical wavelengths, 
its nucleus is made almost completely invisible by
its dust lane, rich in molecular and ionized gas, which obscures
the inner half kiloparsec of the galaxy.  A comprehensive recent
review on the properties of the galaxy and of its AGN is presented
by \citet{israel}.

The kinematics of the ionized gas was mapped in \HA\ by 
\citet{bta87} and subsequently modeled by
\citet[hereafter NBT92]{nbt92} who showed that the dust
lane can be kinematically and morphologically explained with a
thin warped disk geometry.  On the assumption of closed circular
orbits, the \HA\ velocity field is consistent with the rotation
curve expected from the galactic potential.  Kinematical mapping
in CO by \citet[hereafter Q92]{q92} confirmed the previous
results, showing that the material observed in CO is dynamically and
geometrically very similar to that seen in \HA.  \citet{q93} demonstrated that the thin warped disk model can
explain the observed near-IR morphology.  \citet{schreier96}
showed that HST R-band imaging polarimetry is also consistent
with dichroic polarization from such a disk.

The first observational evidence for a large mass concentration at
the nucleus of Centaurus A came from CO kinematical observations
of the nuclear region; these suggested a compact, $100-300\PC$-scale
circumnuclear ring enclosing a 
mass of $\sim\ten{9}\Mo$ \citep{israel90,israel91,ryd93,hawa93}.
ISO imaging by \citet{mira99} provided morphological evidence
that the molecular material within the central 2\KPC\ might be in the form of a 
gaseous bar. \PA\ narrow band images obtained with NICMOS on the HST 
\citep[hereafter Paper I]{schreier98}, revealed an
elongated ($\sim 2\arcsec\times1\arcsec$) emission line region centered on 
the nucleus; this was interpreted as an extended
accretion disk around the AGN.
\citet[Paper II]{marconi00} presented a detailed study of the
photometric structure of the galaxy interior to the dust lane using
NICMOS and WFPC2 which led to a black hole mass estimate of
$\ten{9}\Mo$, using BH growth models by \citet{marel99}.
To confirm 
the presence of a black hole, and to better measure its mass, requires high 
spatial and spectral resolution IR spectroscopy, as is now becoming 
possible with the new generation of 8m-class
telescopes, with state of the art infrared arrays.

In this paper, we report the results of a high spatial resolution
near-IR spectroscopic study of both the ionized and molecular gas
in the central region of the galaxy.  Our data, obtained with
ISAAC at the ESO {\it Very Large Telescope}, allow us to study
the kinematics of the circumnuclear disk discovered in Paper I on 
sub-arcsecond scales and to obtain a dynamical mass estimate for
the central black hole.  In Section \ref{sec:obs} we discuss the
spectroscopic observations and data reduction. The results
themselves are described in Section \ref{sec:res}, which also
discusses the difference in the kinematics between the ionized
and the molecular gas, and compares our results with those obtained
previously at other wavelengths.  Section \ref{sec:fit} presents
our kinematic modeling of the ionized gas velocity field of the
nuclear disk. In Section \ref{sec:discuss}, we summarize the evidence for a
supermassive BH at the nucleus of Cen A and the consequences for
understanding AGNs and the BH-mass vs bulge-mass relationship.
In particular, in Section \ref{sec:gasbar} we discuss the kinematics of the
molecular gas and provide direct evidence for the kinematic
signature of a weak gaseous bar.  Our conclusions are presented
in Section \ref{sec:summary}.

Throughout the paper, we assume a distance to Cen A of 3.5\MPC\
\citep{tonry91,hui,soria96} and an average heliocentric systemic velocity  of
$V_{Hel}=543\pm 2\KM$ \citep{israel}.

\section{\label{sec:obs} Observations and Data Reduction}

Near infrared observations were performed in May, June and July 1999, with
ISAAC mounted at the {\it Antu} unit (UT1) of the VLT telescope.  ISAAC
\citep{isaac} is a near-IR imager and
spectrograph equipped with a 1024x1024 Rockwell detector whose pixels subtend
0\farcs147 on the sky.  The observations consisted of medium resolution spectra
obtained with the 0\farcs3 slit and the grating centered at $\lambda 1.27\MIC$
(J) and $\lambda 2.15\MIC$ (K).  The dispersion was 0.6 \AA/pix and 1.2 \AA/pix
respectively, yielding a resolving power of 10000 in J and 9000 in K.

Seeing during observations (performed in service mode) was in the range
$0\farcs4-0\farcs6$, with photometric conditions.  At a given position angle of
the slit, the observation procedure was to obtain an acquisition image in K and
then center the slit on the prominent nuclear peak (Paper I). For off nuclear
positions a small telescope movement ($\pm 0\farcs3$ perpendicular to the slit)
was subsequently performed.  After positioning the slit, observations consisted
of several pairs of exposures in which the object was moved to two different
positions along the slit (which we call A and B), in order to perform sky
subtraction. This led to an effective slit length of 45\arcsec. The on-chip
integration time was 180s resulting in a total on-source integration time of
36m at each slit position.  The slit positions are summarized in Fig.
\ref{fig:slitpos} and the observation log is given in Tab. \ref{tab:log}.  The
orientation of the NUC-1 slit was chosen to align with the major axis of the
\PA\ disk. To give a clearer picture of any kinematical signature of the
molecular bar, the NUC-2 slit was oriented along the bar major axis.

Data reduction was standard for near-ir spectroscopy and performed using
IRAF\footnote{IRAF is made available to the astronomical community by the
National Optical Astronomy Observatories, which are operated by AURA, Inc.,
under contract with the U.S. National Science Foundation.}.  In each A or B
frame, we first removed the electronic ghost using task {\it is\_ghost} in the
ESO ECLIPSE package \citep{eclipse}.
The frames were then flat fielded with differential
spectroscopic dome exposures and sky subtraction was performed for each
couple of frames (A-B).  Corrections for geometrical distortion of the slit
direction and wavelength calibration were performed using OH sky line present
on the detector \citep{oliva92}.
We verified that the rectification
was performed to better than 0.1--0.2pix on the detector parts where the object
is located ($\sim 300$pix along the slit).  The relative accuracy of wavelength
calibration was better than 0.2\AA\ in K and J.  After the 2 dimensional
rectification procedure, we obtained two single spectra from each A-B frame and
removed residual sky emission by fitting and subtracting a linear polynomium
along the slit.  Correction for the instrumental response and telluric
absorption were performed by using the O6/O7 star HD113754, observed each night
at the beginning and at the end of the observations.  The intrinsic spectral
shape of the standard star was removed by multiplying by $\lambda^{-4}$ which
is typical of O5-9 stars in the near-ir.

The final 2d spectrum at each slit position was then obtained  by averaging
together all single frames.  First, the frames were aligned along the slit via
cross-correlation and shifts along the dispersion direction (due to grating
movements between different instrument setups) where corrected using the OH sky
lines present in each frame. With this correction the zero point of the
wavelength scale has an accuracy of 0.4\AA\ corresponding to a velocity error
of 10\KM\S\1\ in J and 6\KM\S\1\ in K.  All frames were then shifted to a
common center and coadded rejecting bad pixels/cosmic rays.

We then selected the spectral regions containing the lines of interest and
subtracted the continuum by a polynomial/spline fit pixel by pixel along the
dispersion direction.  The continuum subtracted lines were fitted, row by row,
along the dispersion direction with Gaussian or Lorentzian functions using the
task LONGSLIT in the TWODSPEC FIGARO package 
\citep{longslit92}. When the SNR was insufficient (faint line) the fitting was
improved by co-adding two or more pixels along the slit direction.  All
velocities where then corrected for Heliocentric motion.  The results of this
analysis are shown in the next section.

\section{\label{sec:res}Results}

\subsection{\label{sec:stability}Accuracy and stability of the slit position}

Before analyzing the results of the observations, we discuss
the accuracy of the target acquisition within the slit and the
stability of the slit position during integration on source.

The target acquisition strategy consisted in centering the slit
on the K band nuclear peak identified on the acquisition image.
Given the strength of the K peak compared to the rest of the
galaxy (cf. Paper I), the nucleus can be located accurately to
better than a small fraction of the slit size ($<0\farcs1$ given
the seeing of the observations).  A possible concern arises from
observing in J following acquisition in K, if differential
atmospheric refraction between J and K can move the nucleus out
of the slit.  We calculate, however, that in all cases differential
atmospheric refraction can move the object less than 0\farcs05\ 
perpendicular to the slit, and thus this effect is negligible.

The accuracy of the slit positioning is checked by comparing the rotation 
curves obtained at the same slit position in independent
observations -- the observations in J at NUC-1 and in K at OFF-2 were 
repeated.  The comparison in shown in Fig.
\ref{fig:accuracy} for \PB\ and \FeII\ at NUC-1 and \HMOL\ at OFF-2. The 
filled squares are the repeated observations, with better
seeing.  It can be seen that the rotation curves are in agreement
within the errors. Also, as expected, those obtained with worse
seeing are smoother. The agreement shows that the procedure
for target acquisition is quite accurate and that the offsets
between repeated observations are much smaller than the slit size.

By comparing the 2d line maps obtained from single A/B frames,
we also verified that any movement of the slit during observations does not 
significantly affect the rotation curves. In Fig. \ref{fig:accuracy2}, we 
show the rotation curves obtained from
\HMOL\ at NUC-1 from each A/B frame.  This particular set of data
represents the worst case: the continuum-subtracted position-velocity maps 
show the largest differences, indicating the greatest
amount of slit movement during the observations.  Allowing for
lower signal-to-noise ratio with respect to the averaged map, all
the A/B frames produce rotation curves which agree within the
errors.  The slit movements were not large enough to significantly
affect the rotation curves, i.e. they were much smaller
than the spatial resolution of the observations (0\farcs5).

\subsection{Nuclear Spectra}

The J and K averaged spectra extracted from $0\farcs6\times 0\farcs3$ 
apertures centered on the nucleus are presented in
Fig. \ref{fig:spec}.  The lines targeted for kinematical studies
are \FeII\,\WL\,12566.8\AA\ and \PB\,\WL\,12818.1\AA, 
\HMOL(1,0)S(1)\,\WL\,21212.5\AA\ and \BG\,\WL\,21655.3\AA\ but we also 
detect other fainter \FeII\ and \HeI\ lines.  The strong K continuum (see 
Paper II for details) leads to a small equivalent
width for \HMOL\ and \BG.

The nuclear spectrum is low excitation, indicated by the $\FeII/\PB\sim 1.7$
\citep[e.g.][]{alonso,moorwood}.  This ratio does not
change with larger apertures which include more emission from the \PA\ disk
detected in Paper I.  Indeed, we showed in Paper II that the
\FeII\WL\,1.64\MIC\ and \PA\ morphologies are similar and that their ratio is
constant over the disk.

The electron density of the emitting gas can be estimated using the ratio
\FeII\,\WL\,12566.8\AA/ \WL\,12942.6\AA\,$=0.15\pm 0.05$. Figure 
\ref{fig:dens}
displays the region in the \Te-\Ne\ plane allowed by the observed ratio. It 
indicates a density in the range 3.5--4.5\CM\3\ (in log),
regardless of electron temperature. The \FeII\ line ratio as a
function of density and temperature was computed using a code kindly provided 
by E. Oliva which solves the
detailed balance equations for a 16-level Fe$^{+}$ ion, using
atomic data from \citet{feiidata}.
In Paper I we estimated the
mass of the ionized gas disk assuming an electron density in the
range \ten{2}-\ten{4}\CM\3.   The current measurement indicates a
mass of ionized gas at the low end of the previous estimate, $\sim
\ten{3}\Mo$.

An emission feature is clearly detected blueward of the brightest \FeII\
line.  Candidates for the identification are \FeII\WL
12521.3\AA, \HeI\WL 12527.4 ($3\,^3S-\, ^3P^o$) or the high ionization 
coronal line \SIX\WL 12523.5\AA\ \citep{clr},
as shown in the figure.  All the lines (except for \SIX) are drawn
at a redshift of $540\KM\S\1$, corresponding to the systemic
velocity.  Clearly, \HeI\ can be excluded since it would require
a blue-shift of $\ge 240\KM\S\1$ which is not observed in other
lines with similar excitation, like \FeII\ and \PB. Moreover, none
of the other \HeI\ lines expected in this spectral range are
detected.  \FeII\WL 12521.3\AA\ can also be excluded since its
ratio relative to \FeII\WL 12566.8\AA\ for $\Ne =  10^4\CM\3$ is at
most 0.01 while the observed feature would imply a ratio $\simeq 0.04$.

If the feature is indeed \SIX, it would require a blue-shift of $\simeq
150\KM\S\1$ with respect to the systemic velocity, as plotted in the figure.
Indeed \SIX\ has been detected in the Circinus galaxy \citep{clr}, NGC 1068
\citep{clr1068} and has been found to be blue-shifted by $\simeq 40\KM\S\1$ and
$\simeq 250\KM\S\1$, respectively \citep[also cf.][]{oliva97}.  Coronal lines
blue-shifted with respect to lower excitation lines are often seen in other
galaxies.  \citet{binette} suggests that this is due to radiative acceleration
of the coronal line clouds powered by the AGN continuum.  Centaurus A has a
very low excitation spectrum and it might be surprising to detect a high
ionization coronal line (to produce the \ion{S}{+8} ion 351\EV\ are required).
However, when comparing data from similar slit sizes, the \SIX/\FeII\ ratios in Circinus and
NGC 1068 (\SIX/\FeII = 0.3 and 0.1, respectively) are much larger, by almost a
factor 10, than in Cen A.  The coronal line emission is usually spatially
unresolved from the ground and concentrated at the nucleus; therefore large
apertures strongly favor \FeII.  Note that the spatial resolution and
signal-to-noise ratio of the spectra presented here is almost unprecedented;
spectra taken with previous instruments and smaller ground-based telescopes
could not detect such faint lines.

Finally, we place an upper limit of $W_\lambda <5$\AA\
to the equivalent width of any broad ($>1000\KM\S\1$) \PB\ component.

\subsection{Gas kinematics}

\begin{figure*}
\centering
\epsfig{figure=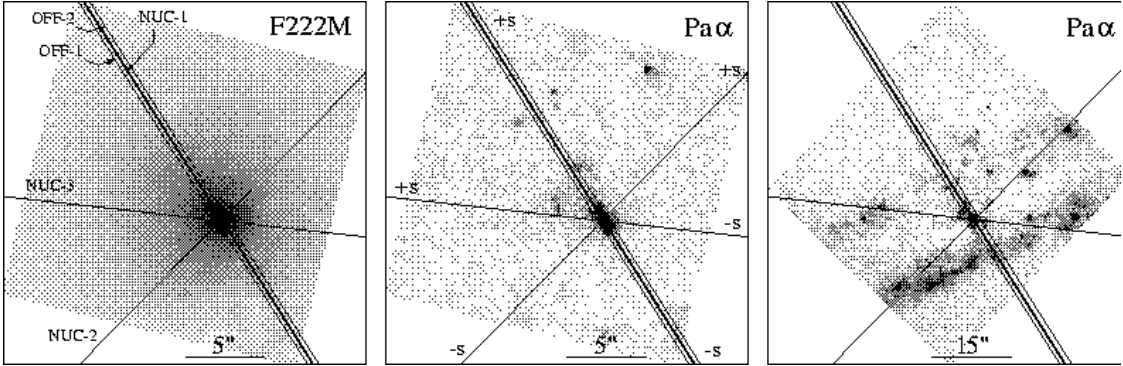,angle=-90,width=0.9\linewidth}
\figcaption{\label{fig:slitpos} Slit positions overlayed, from left to right,
on the F222M and \PA\ NICMOS 2 images (Schreier et al. 1998) and on
the NICMOS 3 \PA\ image by Marconi et al. (2000). The "NUC" and "OFF"
labels indicate the slit positions described in Table 1.
"s+" and "s-" indicate the sign of the position along the slit used in 
the following figures. North is up and East is left.}
\end{figure*}

\begin{figure*}
\centerline{\epsfig{file=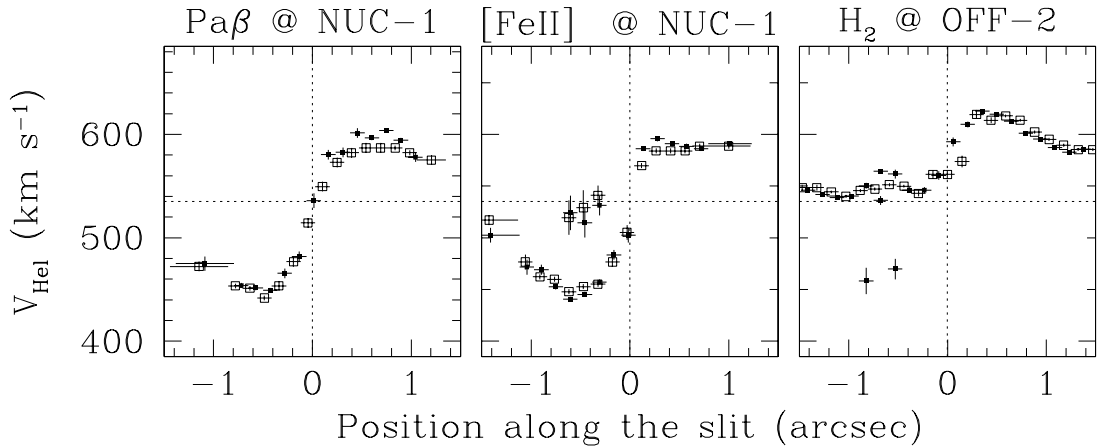,angle=-90,width=\linewidth}}
\caption{\label{fig:accuracy} Comparison of the rotation curves
obtained at different times on the same position. The filled squares
are the velocities obtained with the best seeing (cf. Tab. 1).}
\end{figure*}

\begin{figure*}
\centerline{\epsfig{file=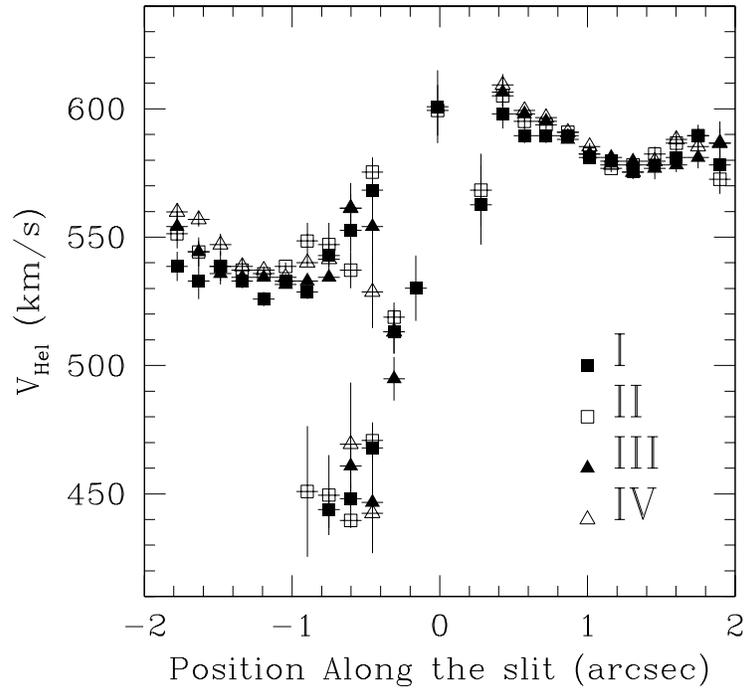,angle=0,width=0.6\linewidth}}
\caption{\label{fig:accuracy2} Comparison of the rotation curves
obtained at NUC-1 from each single \HMOL\ observation. 
All velocities agree within the errors.}
\end{figure*}

\begin{figure*}
\centerline{
\epsfig{figure=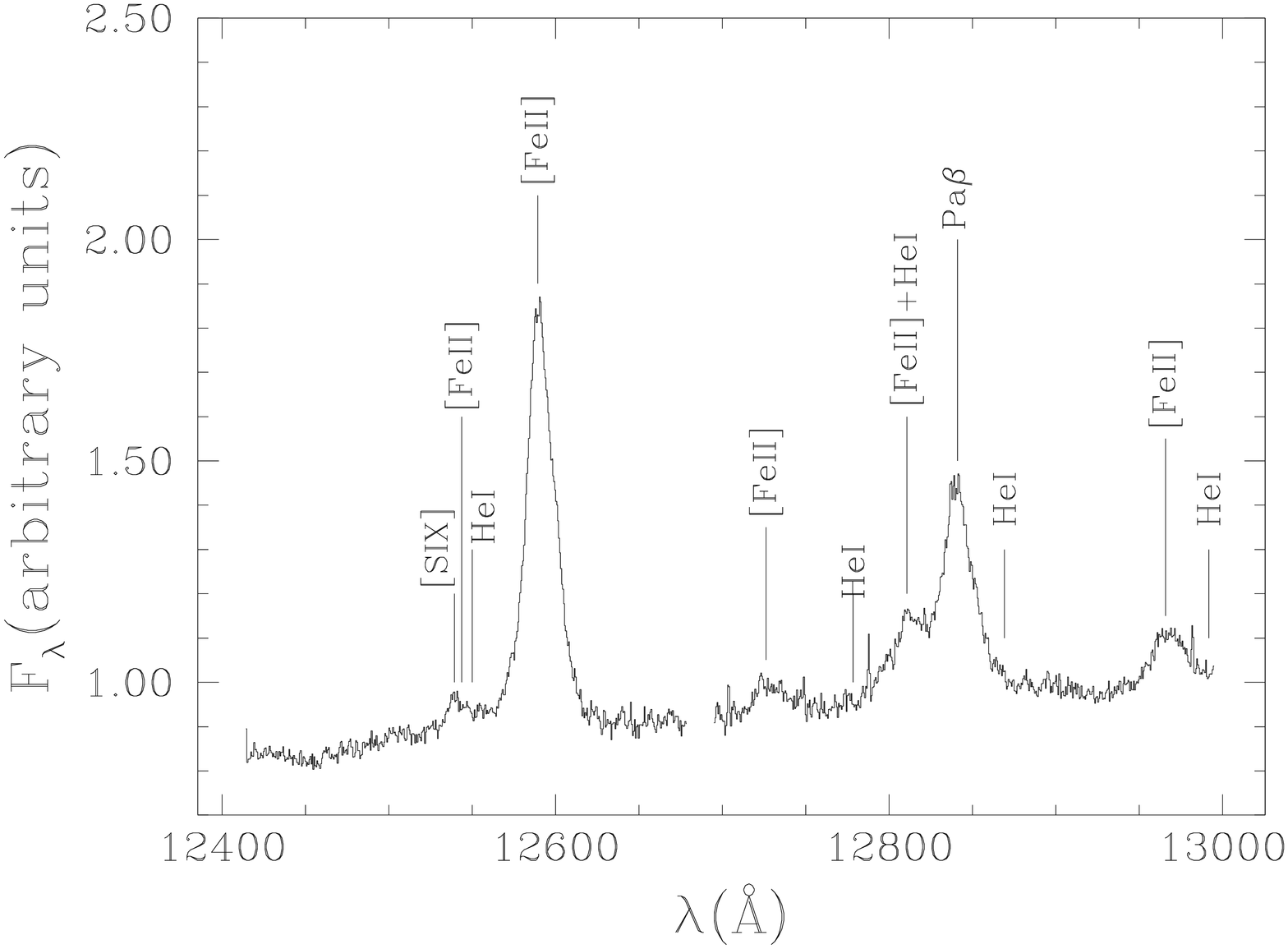,angle=-90,width=0.7\linewidth}}
\centerline{
\epsfig{figure=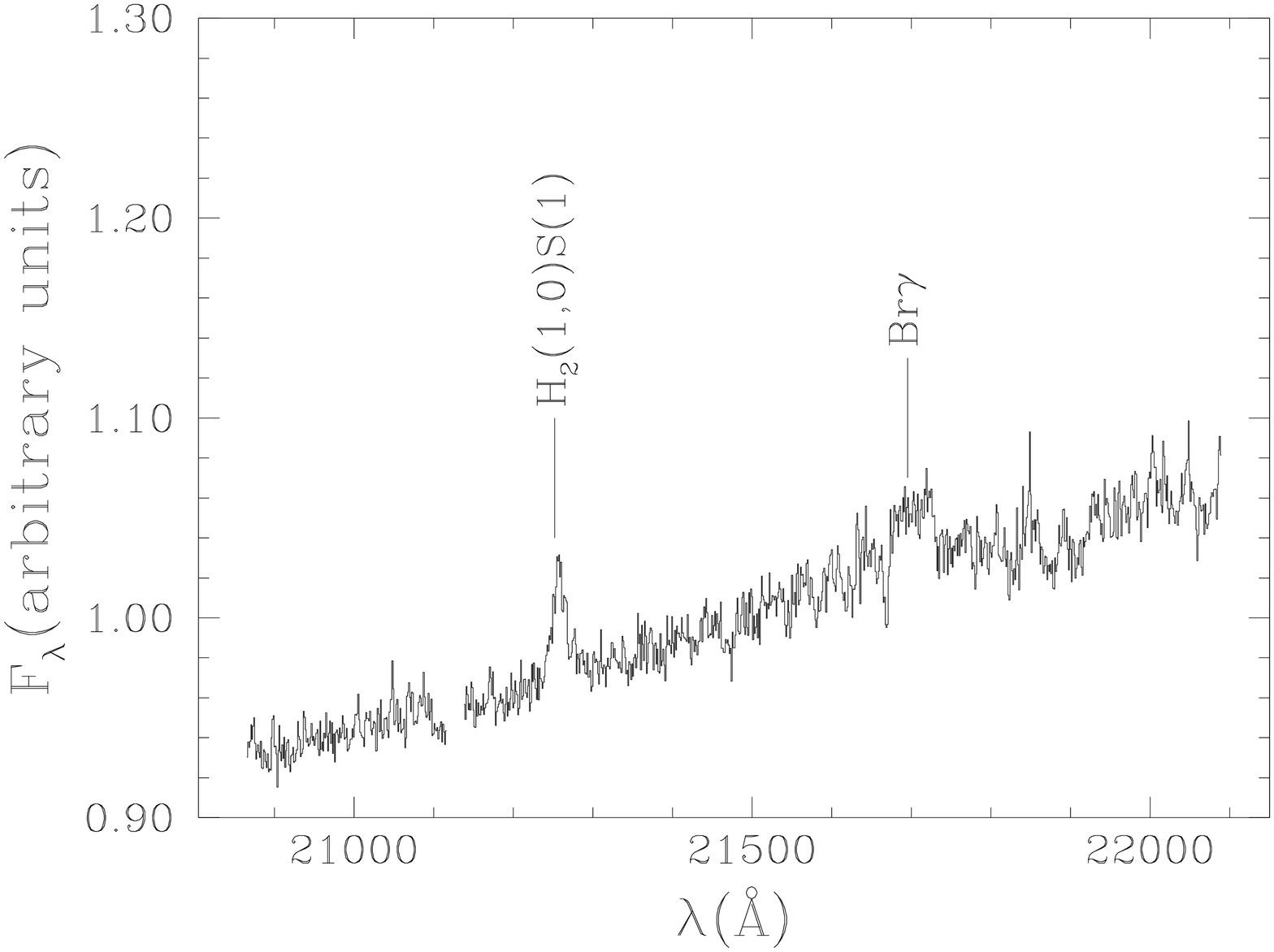,angle=-90,width=0.7\linewidth}}
\caption{\label{fig:spec} Averaged spectra extracted from 
$0\farcs6\times 0\farcs3$ apertures centered on the nucleus.
The X axis is the observed wavelength, the Y axis is flux per
wavelength band in arbitrary units. Lines of interest are marked at the
wavelength corresponding to a redshift of 540\KM\S\1. See text for more
details.}
\end{figure*}

\begin{figure}
\centerline{
\epsfig{figure=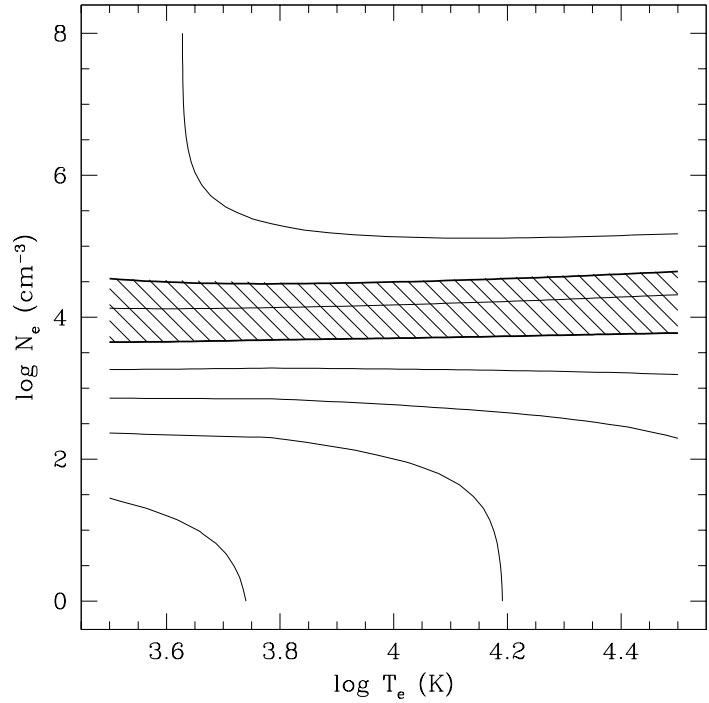,angle=0,width=0.6\linewidth}}
\caption{\label{fig:dens} 
\FeII\,\WL\,12566.8\AA/\WL\,12942.6\AA\ ratio as a function of
electron density (\Ne) and temperature (\Te). The thin lines are the loci
of constant ratio whose represented values
range from -1.6 (in log) in the lower left
corner to -0.6 in the upper right corner with step of 0.2dex.
The thick lines are the lower and upper limit of the observed value
($0.15\pm 0.05$, i.e. -1.0 and -0.7 in log) and the dashed area
shows the permitted values for \Ne\ and \Te. Note that the electron
density is $\log\Ne\sim 3.5-4.5\CM\3$ regardless of temperature.}
\end{figure}

\begin{figure*}
\begin{center}
\epsfig{file=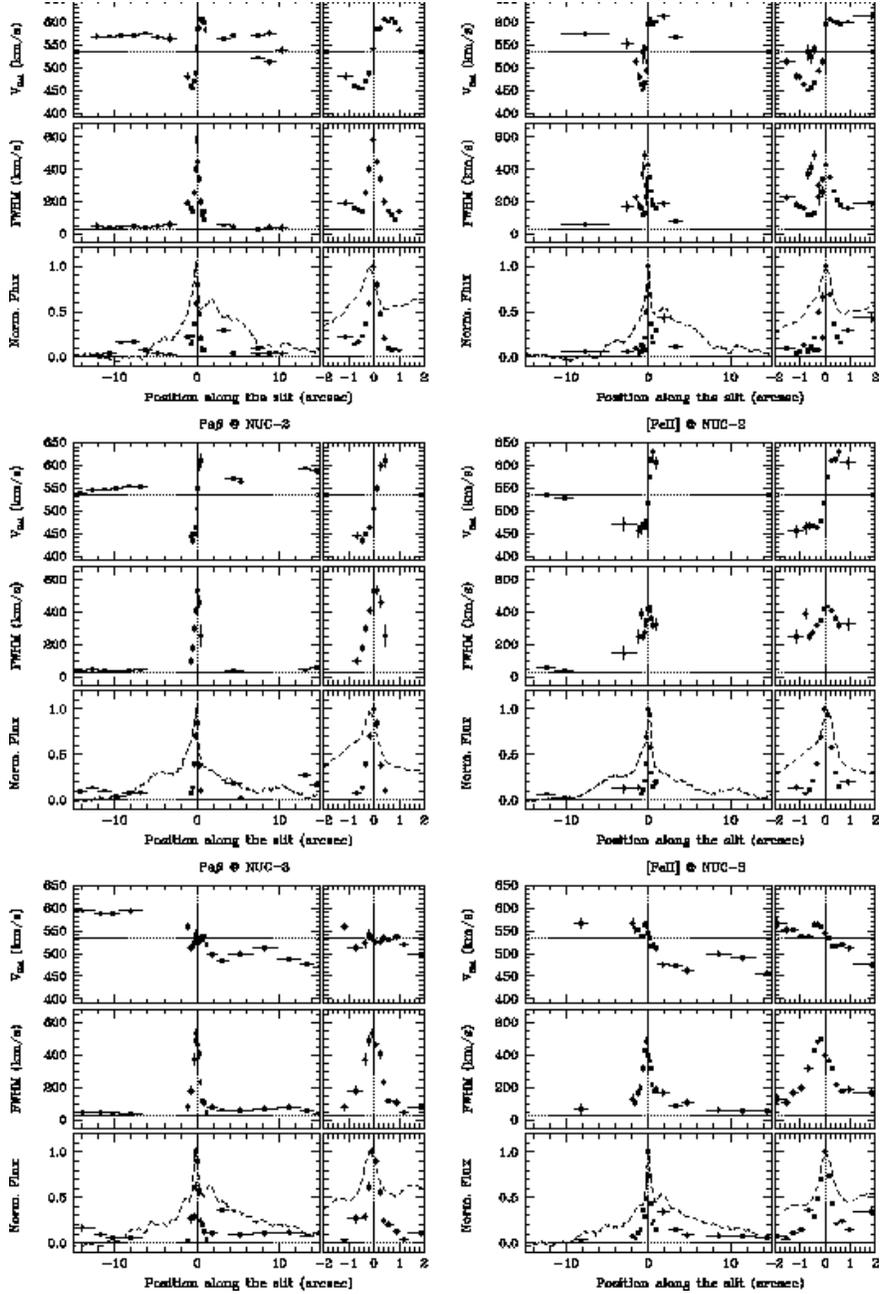,angle=0,width=0.7\linewidth}
\end{center}
\caption{\label{fig:pbferot} Heliocentric velocities (\KM\S\1),
FWHM (\KM\S\1) and fluxes (arbitrary units)
plotted as a function of the position along the slit (arcsec) 
for \PB\ and \FeII. The dotted lines are drawn for
$s=0$ and $V_\mathrm{Hel}=535\KM\S\1$. The dashed lines in the "flux"
panels represent the continuum flux along the slit.}
\end{figure*}

\begin{figure*}
\begin{center}
\epsfig{file=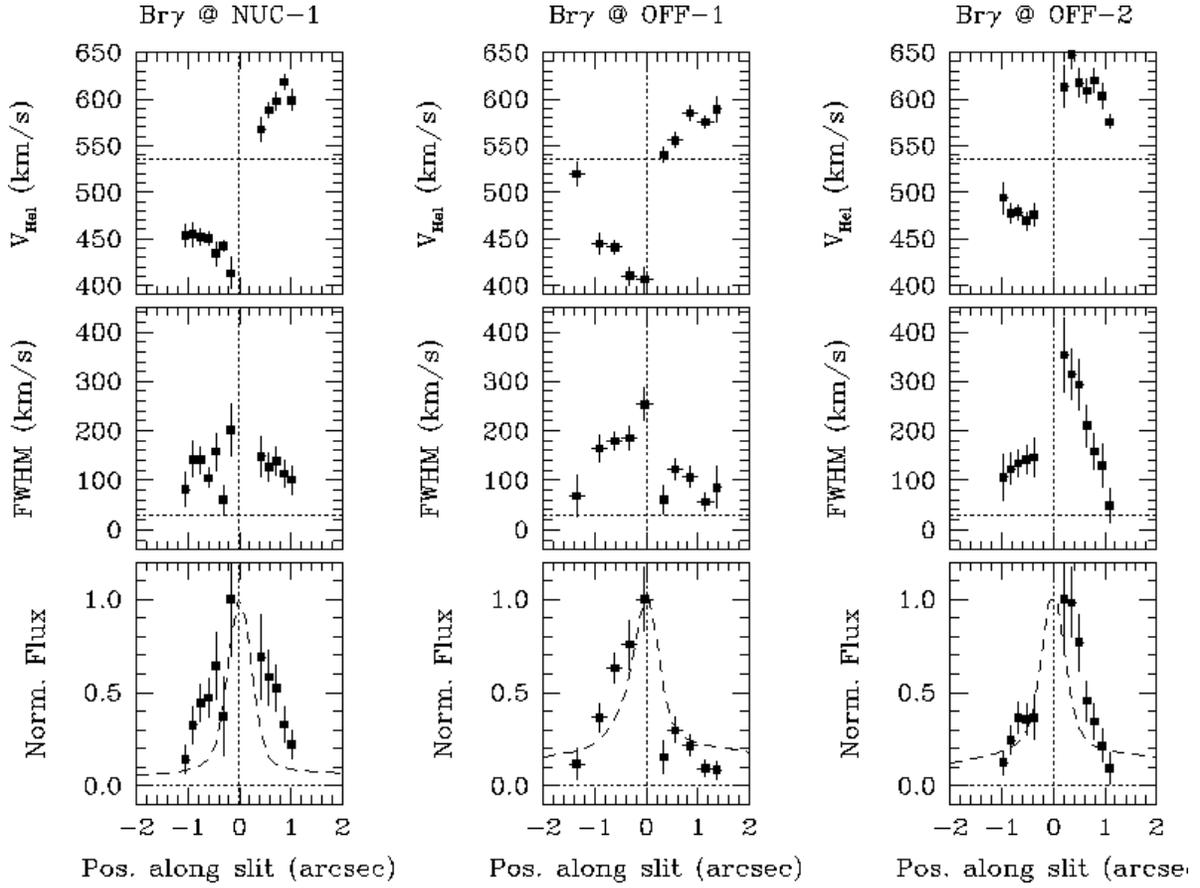,angle=-90,width=0.95\linewidth} \\
\end{center}
\caption{\label{fig:bgrot} \BG\ plots in the inner 2\arcsec\ 
(same notation as in Fig. 6).}
\end{figure*}

\begin{figure*}
\begin{center}
\psfig{file=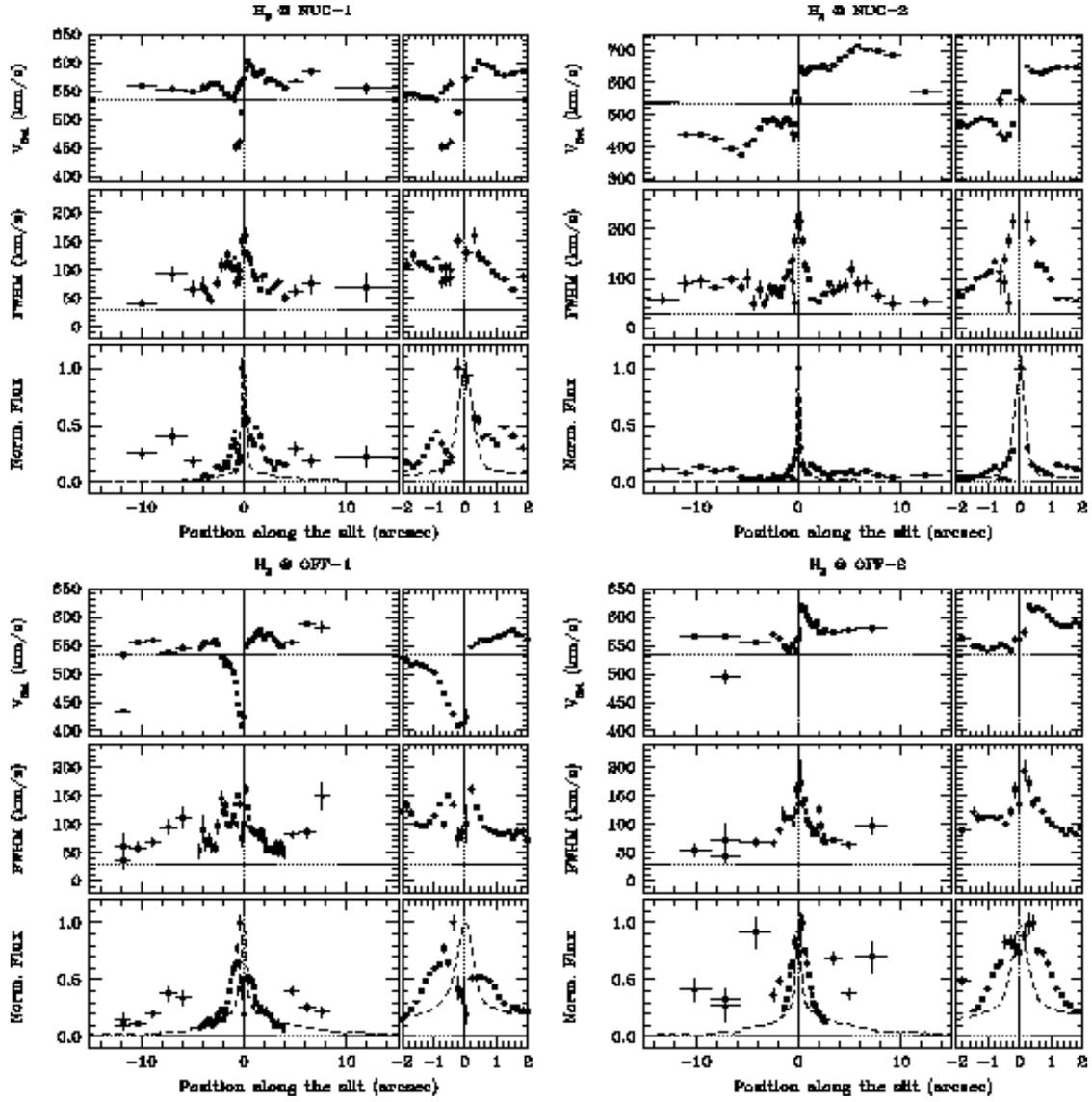,angle=0,width=0.9\linewidth} 
\end{center}
\caption{\label{fig:h2rot} \HMOL\ plots (same notation and layout as in Fig.
6).}
\end{figure*}

Heliocentric velocities, FWHM ({\it Full Width at Half Maximum})
and fluxes measured along each of the slits are shown in Fig. 
\ref{fig:pbferot}, \ref{fig:bgrot} and \ref{fig:h2rot}.  The corresponding 
slit positions are described in Table \ref{tab:log} and shown in Fig. 
\ref{fig:slitpos}.  In the rest of the paper, we will refer to $s$ as the 
coordinate describing the position along the slit. In $K$ the zero point of 
$s$ is defined by the position of the near-IR peak which marks the nucleus 
of the galaxy (Paper I and II).  In $J$ the zero point is also located by 
fitting the rotation curve.  The $+$ or $-$ sign of $s$ is also indicated 
in Fig. \ref{fig:slitpos} to help associate velocities with locations in 
the galaxy.

The ionized gas rotation curves can be decomposed into two distinct
velocity systems. The first is an {\it extended} component 
($|s|>2\arcsec$), characterized by small velocity gradients and a FWHM 
close to the instrumental resolution (30\KM\S\1\ in J and 33\KM\S\1\ in K). 
The second velocity system is confined within
the central $|s|<2\arcsec$ and is characterized by large velocity 
amplitudes and gradients, FWHMs which rise sharply close to $s=0$ and peak 
at $\sim600\KM\S\1$, and by a flux much larger than that
of the extended region.  We refer to this as the {\it nuclear}
component.  These components can be traced also in the molecular
gas but, in addition, a further extended kinematical component is seen in 
\HMOL\ at NUC-2, as shown in Fig. \ref{fig:h2rot} and \ref{fig:greyvel}; 
this component is not present in the ionized gas.

\subsubsection{The extended component}

The velocity field of the extended component of the ionized gas
(cf. at all three slit positions of the \PB\ data) has small
gradients and a quite regular structure which can be followed
out to $\sim 15\arcsec$ from the nucleus.

Our data can be compared with the analysis by NBT92 performed
using optical lines.  Their data, covering most of the dust lane,
are well accounted for by a model in which the gas is distributed
in an optically thin warped disk. Using the full two dimensional
information, NBT92 determined the kinematic line of nodes 
for this ionized gas disk which is found to bend, following the
warping of the dust lane. In the inner 15\arcsec\ (the region
covered by our data) the kinematic line of nodes has a PA $\sim 90^\circ$.  This
is consistent with the fact that the velocity gradient in our
data increases from PA $33^\circ$ to PA $-45^\circ$, reaching
a maximum value of 4\KM\S\1/arcsec at PA $84^\circ$.  This value agrees 
well with the $\sim 3.7\KM\S\1$/arcsec that we derive from
Fig. 7 of NBT92.  At the distance of Cen A (3.5Mpc), it corresponds
to 235 \KM\S\1\KPC\1.  Note that, NBT92 did not find high velocity
gas or line widths larger than 40\KM\S\1\ at the nucleus of Cen A.
This is very likely due to reddening within the inner 2\arcsec\,
where any nuclear \HA+\NII\ emission can be completely hidden
($\AV>5$mag, see Paper II).

The kinematics of the extended gas component can also be studied
from the molecular gas as traced by the \HMOL\ line.  While at
NUC-1 the \PB\ and \HMOL\ velocity curves are very similar, at
NUC-2, the \HMOL\ velocity shows a sharp linear rise between $3\arcsec<s< 
6\arcsec$, with a peak to peak amplitude of
$\sim 300\KM\S\1$ at $\pm6\arcsec$, and then a smooth decrease
out to $10\arcsec$ where the velocity drops to its systemic value.
It traces the kinematics of the molecular gas alone, as it is not
observed in \PB\ and \FeII.  This velocity structure is symmetric with 
respect to the zero point (i.e. $V=V_{sys}$ and $s=0$) and is
suggestive of the rotation curve of a ring, with an inner "solid body 
rotation" and an outer keplerian decrease.  The origin of
this kinematical feature will be discussed in Section 5.3.

\subsubsection{The nuclear component}

At the positions NUC-1 and 2, the ionized gas shows a well defined
{\it S}-shaped velocity curve with peak to peak amplitudes up to
$\sim 200\KM\S\1$.  Velocities from different lines in the inner
2\arcsec\ are compared in Fig. \ref{fig:compare}. The \PB, \FeII\
and \BG\ data agree remarkably well at all slit positions.  There
is a region of small discrepancy at 0\farcs5 seen in the NUC-2 spectrum 
probably caused by contamination from the extended component as discussed 
below.  The \BG\ equivalent width drops dramatically within the central 
0\farcs5 due to the strong nuclear continuum making velocity measurements 
impossible with the present data.

These velocity curves are strongly suggestive of gas rotation within a thin 
disk.  This interpretation is supported by comparing the velocity curves at 
position angles 33\arcdeg and -45\arcdeg\ with that at PA 84\arcdeg. The 
latter has a very small amplitude as expected if that position angle for 
the slit is almost perpendicular to the line of nodes of the disk which, 
therefore, should be oriented close to North-South.  At all positions, we 
find a sharp increase in FWHM near the nucleus, the expected consequence of 
unresolved rotation close to the nucleus: it is not possible to trace the 
rotation curve within a spatial resolution element from
the nucleus, and the mixing of high velocity components broadens the
line profile.  We will return to this in Sec. 4.1.

The \HMOL\ rotation curves in the inner 2\arcsec\ are also compared with 
those of the ionized gas in Fig. \ref{fig:compare}. The amplitude of the 
\HMOL\ rotation curve, particularly at NUC-1, is smaller than those of \PB, 
\FeII\ and \BG.  Note, however, the double \HMOL\ line profiles between 
$-1\arcsec<s<-0\farcs4$ (Fig. \ref{fig:h2double}). When accurate deblending 
is possible, the blueshifted velocity component agrees with the velocities 
measured from the ionized gas (Fig. \ref{fig:compare}).  In the
adjacent points, where deblending is impossible, the width of the lines is 
larger ($\sim 100\KM\S$) and the measured velocity is an intensity weighted 
average of the two components.  The "contamination" from the extended 
component is particularly evident for $s<0\arcsec$, i.e.  south of the 
nucleus (cf. Fig.
\ref{fig:slitpos}), at NUC-1 and OFF-2. It is also responsible for the 
sharp decrease in \HMOL\ velocity observed at OFF-1 below $-0\farcs6$.  The 
discrepant \FeII\ velocities at NUC-2 are also likely caused by mixing of 
these 2 velocity components and the increased \FeII\ line widths at this 
location support this interpretation.

The high velocity components are thus 
present also in the molecular lines but they are partly hidden by the 
superposition with the extended component. However, when deblending is 
possible, velocities from the rotating disk can still be measured and they 
are in agreement with those from the ionized gas. This "contamination" is 
likely to be present also in the other lines but their extended components 
are much weaker with respect to the nuclear ones (see the line fluxes along 
the slit in the figures) and do not affect the velocity determination in 
the inner 2\arcsec.

\section{\label{sec:fit}Kinematical Models of the nuclear rotation curves}

The nuclear velocity curves described in the previous section clearly 
suggest gas rotating in a disk around a compact mass. We now investigate if 
they are indeed quantitatively consistent with this interpretation.  Under 
the hypothesis of a thin gaseous disk circularly rotating in the 
gravitational potential of a point mass $M$, the rotation curve at a given 
slit position can be written as \citep{m87bh}
\begin{equation} V = V_{sys} - (GM)^{0.5}\sin i\frac{X}{R^{1.5}}
\end{equation} where $R$ is the deprojected radial distance, $X$ is the 
distance from the disk minor axis and $i$ is the disk 
inclination with respect to the line of sight ($i=0$ is the face-on case). 
The geometry and definitions are also summarized in Fig. \ref{fig:diskgeo} 
and in the Appendix.

This formula provides the velocity along a thin slit but is not immediately 
applicable since the finite spatial resolution of the observations smears 
the velocity cusps of the rotation curve. Moreover, the measured velocities 
are intensity weighted averages over apertures determined by the slit size 
and the
number of pixels integrated along the slit.  To simulate the effects of 
seeing, we convolve with a gaussian PSF ({\it Point Spread Function})
with FWHM equal to the seeing of the observations ($0\farcs5$).
We then integrate over the slit size to derive the expected velocity 
$V_{psf}$ as a function of the slit coordinate $s$.

The observations provide us with several datasets, each from a given line 
(\PB, \FeII\, \HMOL\ or \BG) at a given slit position (NUC-1, NUC-2, NUC-3, 
OFF-1 and OFF-2).  Each data set $k$ comprises $n_k$ points and each point 
is characterized by its position along the slit $x_i$ (in pixels, i.e. 
detector coordinates) and its velocity $V_i$ (in \KM\S\1) derived from the 
line fitting procedure.  Three free parameters are associated with each 
dataset:
\begin{itemize}
\item \Xok, the pixel at which $s=0$;
\item \B$_k$,
the distance of the center of the slit from the nucleus;
\item \Th$_k$, the angle between the slit and the line of nodes of the disk;
\end{itemize}

\Xok\ is fixed at the position of the continuum peak.  In the K
band, the continuum peak undoubtedly identifies the location of the 
nucleus, given its strength relative to the surrounding emission from the 
galaxy. In the J band, however, the continuum peak has an intensity 
comparable with that of the extended emission: any asymmetry in the galaxy 
emission (due, e.g., to an extinction gradient) could shift the peak of the 
continuum with respect to the position of the nucleus. Therefore \Xo $_k$ 
must be allowed
to vary with respect to the position of the flux peak by the same amount 
for lines from the same spectrum.  Once \Xok\ and the pixel size are set, 
the coordinate $s$ along the slit is given by 
$s_i=0\farcs147\times(x_i-\Xok)$.

For the nuclear positions NUC-1,2 and 3, \B$_k$ is set to 0, given that the 
acquisition procedure and the telescope guiding accuracy is
well below 0\farcs1. For the off-nuclear position, \B$_k = \pm 0\farcs3$.

Since the relative orientation of each pair of slit positions is fixed by 
the observational strategy, all values of $\Th_k$ are determined by a 
single value \Th, the angle between the line of nodes and the North-South 
direction.

The "global" free parameters are:
\begin{itemize}
\item \I, the inclination of the disk;
\item \M, the black hole mass;
\item \Vs, the systemic velocity.
\end{itemize}

We fit the free parameters by minimizing $\Chi$:
\begin{equation}
\label{eq:chisq}
\Chi = \frac{1}{\sum_{k=1}^{N}n_k-f} \sum_{k=1}^{N}\sum_{j=1}^{n_k}
\frac{\left(V_j-V_{psf}(x_j; \Xok, \Th, \I, \M, 
\Vs)\right)^2} {\Delta V_j^2}
\end{equation}
where N is the total number of datasets and $f$ is the number
of free parameters.

Since, as shown in the previous section, the \HMOL\ rotation curves of the 
nuclear component are contaminated by the extended component, and the 
signal-to-noise and spatial sampling of the \BG\ data is poorer than that 
of \FeII\ and \PB, we fit only the latter two lines. The \HMOL\ and \BG\ 
data are used only as a consistency check after performing the fit.  The 
velocity points excluded from the fit are drawn in grey color in Fig. 
\ref{fig:compare}.

The best fits for fixed values of the disk inclination are
shown in Figs. \ref{fig:fit} and \ref{fig:fit2}; the resulting parameters 
are given in Tab. \ref{tab:fit}.
The agreement between the data and the models is good 
for all three slit positions at disk inclinations below 60\arcdeg.  The 
residuals do not show any systematic trend and their rms is typically $\sim 
10$ \KM\S\1.

Although the best fit models trace the data well, the minimum \Chi\ values 
are far larger than expected for a good fit.  This discrepancy probably 
arises from the fitted data points not being statistically independent. 
From a visual inspection of the fits, we
estimate typical uncertainties on fit parameters as $\Delta\Th\pm 
5\arcdeg$, $\Delta\Vs\pm 5\KM\S\1$, and $\Delta\M/\M\pm 20\%$.

We assess the validity of our results by comparing the model
rotation curves with the observed \HMOL\ and \BG\ curves (see Fig.
\ref{fig:fith2bg}).  In particular, the asymmetric behaviour of the 
off-nuclear rotation curves measured from the \HMOL\ and \BG\ lines, 
typical of the impact parameter being not zero, is naturally accounted for.

We examined the effects of varying the disk inclination on the 
determination of the value of the central mass; these two parameters are 
strongly coupled, since the amplitude of the rotation curve is proportional 
to $\M^{0.5}(\sin i)$.  A weaker dependence on \I\
alone is also present, as the deprojected radius $R$ depends on $\cos\I$ 
(see the Appendix).  We find that the signal-to-noise of our data is 
sufficient to reject all values of \I\ ($> 60\arcdeg$), since the model is 
not able to reproduce the observed velocity amplitude at NUC-1 and NUC-2 at 
the same time. For such large values of inclination, varying the angle with 
the disk line of nodes
maps very different radial distances $R$ into similar slit 
coordinates $s$, thus requiring different masses to reproduce a given 
velocity.  Conversely, all values of inclination $\I\lesssim 60\arcdeg$ 
produce acceptable fits.  The minimum dynamical mass required by the data 
is
$\sim 9\xten{7}\Mo$; this estimate obviously increases with  decreasing 
disk inclination.

It is clear from inspection of Tab. 2 that regardless of inclination, the 
systemic velocity is well determined at $\Vs=532\pm 5 \KM\S\1$,
in agreement with 
published values. Similarly, the position angle of the line of nodes is 
well constrained, in the range $-10\arcdeg\div -14\arcdeg$.
Note that the kinematic line of nodes (PA=-14\arcdeg) determined from our 
velocity curves differs from the axis of elongation of the \PA\ disk 
(PA=33\arcdeg) seen with NICMOS.  This can be explained if only a part
of the disk is ionized and  
lies within the ionization cone (the jet axis projected on the sky has PA 
$\simeq 55\arcdeg$); the ionized portion is then visible in \PB\ and \FeII. 
Another possibility is that the disk density, and thus the emission
measure, is very low outside of the region observed with NICMOS in \PA. 

\subsection{The nuclear line profiles}

The line flux profile along the slit in the inner $|s|<0\farcs5$ is 
strongly peaked and unresolved; its FWHM is comparable to the seeing ($\sim 
0\farcs5$) and is the same regardless of slit orientation. We conclude that 
the line luminosity profile along the slit has a
FWHM less than 0\farcs1; otherwise it would appear spatially resolved. The 
actual surface brightness distribution of the line around the nucleus is 
unknown. If we assume a double exponential profile extending down to a
minimum radius $r_h$ 
\begin{equation}
 I(r) = \left\{
	\begin{array}{ll}
	I_0\,\mbox{e}^{-\frac{r}{r_0}} +
        I_1\,\mbox{e}^{-\frac{r}{r_1}}	& r\ge r_h \\
	0	& r<r_h
	\end{array}
	\right.
\end{equation}
where $r_0$ and $r_1$ are the scale radii,
the values $r_0=0\farcs08$, $r_1=1\arcsec$, $r_h=0\farcs08$
and $I_1/I_0=0.015$ match the line flux distributions along the slit. 

Performing the fit with an intensity profile does
not significantly changes the results as shown in Tab. \ref{tab:fit} and
parameters relative uncertainties are also the same as before.
A fit example is shown in Fig.
\ref{fig:fitint}.  All the parameters agree within uncertainties with the
estimates performed with a constant surface brightness. The only significant
difference is a slightly worse value of \Chi.  Note that, though within the
20\%\ uncertainty, the masses are slightly lower than in the case with 
constant
light distribution. The reason is that the adopted light distribution places
more weighting to the points closer to the nucleus and a given velocity
amplitude can be reproduced with a smaller mass.  However, we have also
verified that surface brightness distributions which are strongly peaked 
toward
the center are excluded because they give too much weight to the high 
velocity
components.  This explains why a "hole" in line emission around the
nucleus is required.

With this assumption on the surface brightness line distribution,
we compute the line profiles expected from a rotating disk and compare them 
with the observed ones.  The detailed formula is
given in the appendix.  Assuming pure gravitational motion of the gas, 
differing line profiles are the consequence of unresolved rotation in the 
inner parts.  Therefore, a crucial test for the keplerian rotation 
assumption is verifying that the observed line widths are consistent with 
unresolved rotation and not the result of non-gravitational motions.

In Figure \ref{fig:prof}, we compare the model profiles with the observations. 
We have assumed intrinsic gaussian line profiles, with FWHM increasing 
linearly from 50\KM\S\1\ at 1\arcsec\ from the nucleus to 250\KM\S\1\ at 
0\farcs1.  The line profiles are computed for \I=45\arcdeg\ with the 
corresponding parameters as in Table \ref{tab:fit}.
There is reasonable agreement 
between the model profiles and the observed data which could be improved 
further by optimizing the intrinsic FWHM of the line profiles.  In Sec.
 \ref{sec:stability} 
we showed that the slit positioning was stable during the observations by 
comparing rotation curves
obtained from independent observations. Though a small movement of the slit 
does not produce any appreciable change in the line centroid, it can change 
the line profiles.  Therefore, some differences between model and observed 
line profiles can be accounted for by small movements of the slit.
Another point to be considered is the dependence of the line profiles on 
the assumed surface brightness distribution.  The most important point is 
that at all slit positions, unresolved rotation can fully account for the 
observed FWHM of the lines and we can  reproduce the observed line 
profiles.  A more detailed analysis of this kind requires higher spatial 
resolution.  Planned HST observations should be characterized by higher 
resolution and better
stability in slit positioning.

\subsection{Stellar contribution to the dynamical mass}

The observed rotation curves require a mass of $\M> 6\xten{7}\Mo$ within 
$R<0\farcs25$, as set by the spatial resolution of our observations. It is 
crucial to determine if this large mass can be accounted for by stars. In 
Paper II, we reconstructed the nuclear light profile in the V band 
(assuming a color-based reddening correction and no color gradients in V-K 
and V-H) and showed that it can be well fitted with a Nuker law with a 
break occurring
at $r_b \sim 4 \arcsec$ (see Fig. 9 in Paper II and Fig. 
\ref{fig:nuker}a).  For the inner $R<10\arcsec$, we used the NICMOS F222M 
image whose isophotes (after reddening correction) are characterized by 
small ellipticities ($<0.1$) and are therefore consistent with a spherical 
stellar system.  When a system is characterized by a spherical light 
density ${\cal J}(R)$ the observed surface brightness is given by
\begin{equation}
\Sigma(r) = 2\int_r^{+\infty} \frac{{\cal J}(R) R dR}{\sqrt{R^2-r^2}}
\end{equation}
where $r$ is the projected radius onto the plane of the sky and
$R$ is the distance from the nucleus.  Once $\Sigma(r)$ is obtained from 
observation, the above equation can be inverted to derive  ${\cal J}(R)$ 
using Abel's formula:
\begin{equation} {\cal J}(R) = -\frac{1}{\pi} \int_R^{+\infty}
\frac{d\Sigma(r)}{dr} \frac{dr}{\sqrt{r^2-R^2}}
\end{equation}
The mass density is then given by $\rho(R) = 4\pi\,
\Upsilon{\cal J}(R)$ where $\Upsilon$ is the mass-to-light ratio, and  the 
mass enclosed within the radius $R$ can be computed.
In Fig. \ref{fig:nuker}a, we plot the reddening corrected K band surface 
brightness points along with the best fit nuker-law profile presented in 
Paper II (solid line). The dashed line represents the nuker law with the 
steepest profile consistent with the data.
In Fig. \ref{fig:nuker}b we show the derived luminosity density (\LKo\ is 
the solar luminosity in the K band) and, finally, in Fig. \ref{fig:nuker}c, 
we show the enclosed mass within a given
radius $R$ for a Mass-To-Light ratio, $\Upsilon=1\Mo/\LKo$.
Note that the steepness of the nuclear cusp is fundamental to estimating 
the enclosed mass at small radii.

Within $R=0\farcs25$, the scale probed by our observations, the Mass-to-Light
ratio required to account for the rotation curve of the ionized disk is
$>20\Mo/\LKo$ ($>6\Mo/\LKo$ with the steepest nuclear cusp consistent with the
data). The observed mass-to-light ratio of elliptical galaxies with total mass
around \ten{11}\Mo\ lies in the range $\Upsilon=0.15-0.5\Mo/\LKo$
\citep{mobasher} and in bulges of early type spiral galaxies is typically
$\Upsilon\simeq0.6\Mo/\LKo$ \citep{moriondo}.  According to stellar population
synthesis models, mass-to-light ratios up to $\Upsilon\sim 1.3$ at $t=15$ Gyr
can be obtained with a Salpeter Initial Mass Function \citep{maraston}.
Though with a very steep IMF the stellar population can reach $\Upsilon\simeq
4\Mo/\LKo$ after 15 Gyr, it is clear that the mass-to-light ratio required by
our observations is incompatible with a "normal" stellar system and suggests
the presence of a {\it dark} mass concentration.
 
The  central mass density is $\simeq 3\xten{5}M_8\, \Mo\PC\3$
($M_8$ is the mass in units of \ten{8}\Mo) and
similarly high mass densities have been observed in a few core-collapsed 
globular clusters \citep{gc}.  We therefore cannot strictly 
rule out that the central dark mass concentration
is an "exotic" object such as a massive cluster of neutron stars or other 
dark objects; an extensive discussion of such possibilities has been given 
by \citet{marel}. However we concur with their general 
conclusion that these alternatives are both implausible and contrived.  We 
believe that the more conservative
explanation for the central mass concentration observed in Centaurus A is 
indeed a supermassive black-hole.

To summarize, the observed stellar light profile cannot explain the 
observed rotation curves for any reasonable choice of M/L, leading us to 
conclude that a dark mass of $\M>6\xten{7}\Mo$ is required.
The lower limit for the Black Hole mass in Cen A is directly constrained by 
the observed rotation curves, but the upper bound is essentially 
unconstrained (e.g., for $\I= 5\arcdeg$, we would obtain a mass value two 
orders of magnitude larger).  Nonetheless, a very small inclination can be 
reasonably excluded by the properties of the radio jet in Cen A, whose 
direction is thought to be very close to the plane of the sky. From VLBI 
observations, \citet{tingay}
derived a value for jet inclination with 
respect to the line of sight of 50\arcdeg--80\arcdeg.
Near-IR polarimetry of Cen A with NICMOS \citep{cap00} indicates 
that emission from the nucleus escapes within a light cone, visible in 
polarized light. This is possible only if the the line of sight is outside 
the cone, whose axis must then be close to the plane of the sky. Although 
there might be a misalignment between the jet and the ionized disk, it 
appears very unlikely that a jet in the plane of the sky could be 
associated with a face-on disk. Excluding only the smallest disk inclinations 
($\I<15\arcdeg$), an upper limit on the BH mass can be set at 
$\M=5\xten{8}\Mo$. We conclude that the black hole mass estimate is 
$\M=2^{+3.0}_{-1.4}\xten{8}\Mo$.

\section{\label{sec:discuss}Discussion}

The observations presented above provide strong evidence for a dark mass 
concentration, very likely a supermassive black hole, at the nucleus of Cen 
A. The BH sphere of influence has an outer radius
\begin{equation}
r_\bullet\simeq \frac{G\M}{\sigma^2} =
	 11\PC \left(\frac{\M}{\ten{8}\Mo}\right)
	 \left(\frac{\sigma}{200\KM\S\1}\right)^{-2}
\end{equation}
where $G$ is the gravitational constant and $\sigma$ is the velocity dispersion
of the stars.  In Cen A, the velocity dispersion has a mean value of $\simeq
120\KM\S\1$ between 40\arcsec\ and 80\arcsec\ from the centre with a nuclear
value of $\sim 150\KM\S\1$ \citep{wilk86}.  With the caveat that extinction
could lead to an underestimate of the nuclear value, the radius of influence of
the BH would then be $r_\bullet\sim 20 M_8 \PC$ i.e. $\sim 1.1\arcsec M_8$.
Indeed the nuclear rotation curves within 2\arcsec\ of the nucleus do not show
any evidence of an extended mass distribution. The radius of influence of the
BH could also be estimated as the radius where the enclosed stellar mass equals
\M.  From Fig. \ref{fig:nuker}c, for $\Upsilon=1 \Mo/\LKo$ this takes place at
$R\sim 1\arcsec$, or 17pc, (for \M=\ten{8}\Mo), consistent with the above
estimate.

The presence of a supermassive BH at the nucleus of Cen A 
has implications for the energetics of the active nucleus and 
for the general BH-mass vs bulge-mass relationship.

\subsection{BH mass and AGN activity}

In the 2-10\KEV\ range, the nucleus of Centaurus A has an unabsorbed X-ray
luminosity of 6\xten{41}\ERG\S\1\, with a power law spectrum typical of normal
AGNs - $\Gamma=1.96\pm0.1$ \citep{turner97}. For a typical quasar,
$L_X(2-10\KEV)/L_\mathrm{bol}\sim 0.03$ \citep{elvis}; assuming that Cen A has
a similar Spectral Energy Distribution (SED), its expected bolometric
luminosity is $\simeq 2\xten{43}\ERG\S\1$, in agreement with other estimates
presented in literature \citep[e.g.][]{kinzer}.  This value indicates that Cen
A is radiating well below its Eddington luminosity as
$L_\mathrm{bol}/L_{Edd}\simeq 2.4\xten{-3} M_8^{-1}$, where $M_8^{-1}$ is the
BH mass in units of \ten{8}\Mo.  This is similar to the case of M87, another
well-studied radio galaxy characterized by a prominent jet and a supermassive
BH \citep[c.f.][]{m87bh}.  The HST survey of nuclei of FR~I radio-galaxies by
\citet{chiab} indicates that sub-Eddington luminosities are typical of this
class of sources to which Cen A belongs.

Luminosities much below the Eddington limit are the result either of a very low
accretion rate or of an inefficient radiation process.  In the latter case it
has been suggested that the accretion is dominated by advection
\citep[Advection Dominated Accretion Flow - ADAF; e.g.][]{narayan} 
in which case the continuum emitted by the AGN lacks
the UV "big-blue bump" emitted by the standard thin accretion
disk \citep[e.g.][]{kup00,dimatteo}.

Using ISO LWS spectra, \citet{unger00} have shown that the total FIR luminosity
of Cen A in a 70\arcsec\ beam centered on the nucleus is
$\LFIR=3.2\xten{9}\Lo$, comparable to the luminosity expected from the AGN,
$\LAGN\simeq5\xten{9}\Lo$, if it has the typical spectrum of a quasar.  If
the latter assumption is true, the sub-Eddington luminosities are the
consequence of a low accretion rate.  However, since FIR emission lines have
excitation and equivalent widths typical of starburst galaxies, \citet{unger00}
suggest that the total FIR luminosity of Cen A is dominated by star formation
instead of AGN activity.  In this case the bolometric correction applied to the
AGN X-ray emission is overestimated by a large factor (a factor of 10 would
imply $\LAGN/\LFIR\simeq 0.15$) suggesting that the big blue bump is weak or
absent as it happens in the ADAF regime.  Indeed, in Paper II we showed that
the extrapolation of the X-ray power law to the optical and near IR can fully
account for the observed nuclear emission.

\subsection{The BH-mass vs bulge-mass relationship}

There is increasing evidence that black holes exist in most or all galaxies,
and it has been suggested that a rough correlation exists between the mass of
the black hole and the mass of the bulge of the host galaxy
\citep[e.g.][]{kr95,rich98}.  In this framework, we can compare the central dark
mass concentration in Cen A with the total mass of the galaxy.

The total mass of Cen A has been derived by photometrical and kinematical 
studies. \citet{hui} and \citet{mathieu}
find $M_\mathrm{sph}=(4\pm1) 
\xten{11}\Mo$. However, in the BH-bulge mass relationships, the spheroid 
mass is usually determined by assuming a mass-to-light ratio from the 
spheroid luminosity in a certain band (usually V or B).  To compare Cen A 
with existing BH-bulge relationships, we thus have to verify that the 
photometry results are compatible with the kinematics.  Indeed, the total V 
mag of Cen A is V=5.75mag (from \citet{dufour}, after correcting for 
the nuclear light profile - Paper II), which corresponds to \MV=-21.95mag. 
Therefore $\log (L/\LVo) = 0.4(-\MV+4.83)=10.7$.  For normal galaxies, 
$M/\LV=5(L/\ten{10}\LVo)^{0.2}$ in solar units, and therefore the expected 
mass of Cen A is $M\sim 3\xten{11}\Mo$ in
good agreement with the kinematical determination.

For Cen A, then:
\begin{equation}
 \frac{\M}{M_\mathrm{sph}}\simeq 2.5\xten{-4} M_8
\end{equation}
where $M_8$ is the BH mass in units of \ten{8}\Mo.  The well known BH-mass vs
bulge-mass relationship \citep{rich98} shows that $\log(\M/M_\mathrm{sph})$ has
a roughly gaussian distribution, with an average of -2.27 (i.e. a 0.005 ratio)
and standard deviation of 0.5.  The expected relationship from Quasar light is
$\simeq 6\xten{-4}$, a factor ten lower.  This apparent discrepancy suggests
that AGN accretion is less efficient than previously thought or that most of
AGN activity is obscured \citep{fabian}.  Note however that the large
observed value of 0.005 is based on the \citet{magorrian} paper where, given
the simplified assumptions on the stellar systems and the low spatial
resolution of the spectroscopic data, BH masses are likely overestimated.
Indeed
Cen A is almost 2.5$\sigma$ (in log) below the average of the Richstone et
al.~and Magorrian et al.~papers while it is in good agreement with the
expectation from Quasar light. The observed $\M/M_\mathrm{sph}$ for Cen A is
also in agreement with the same ratio determined by \citet{wandel} for Seyfert
1 galaxies from reverberation mapping.


\subsection{\label{sec:gasbar}The \HMOL\ "ring"}

\begin{figure*}
\begin{center}
\psfig{file=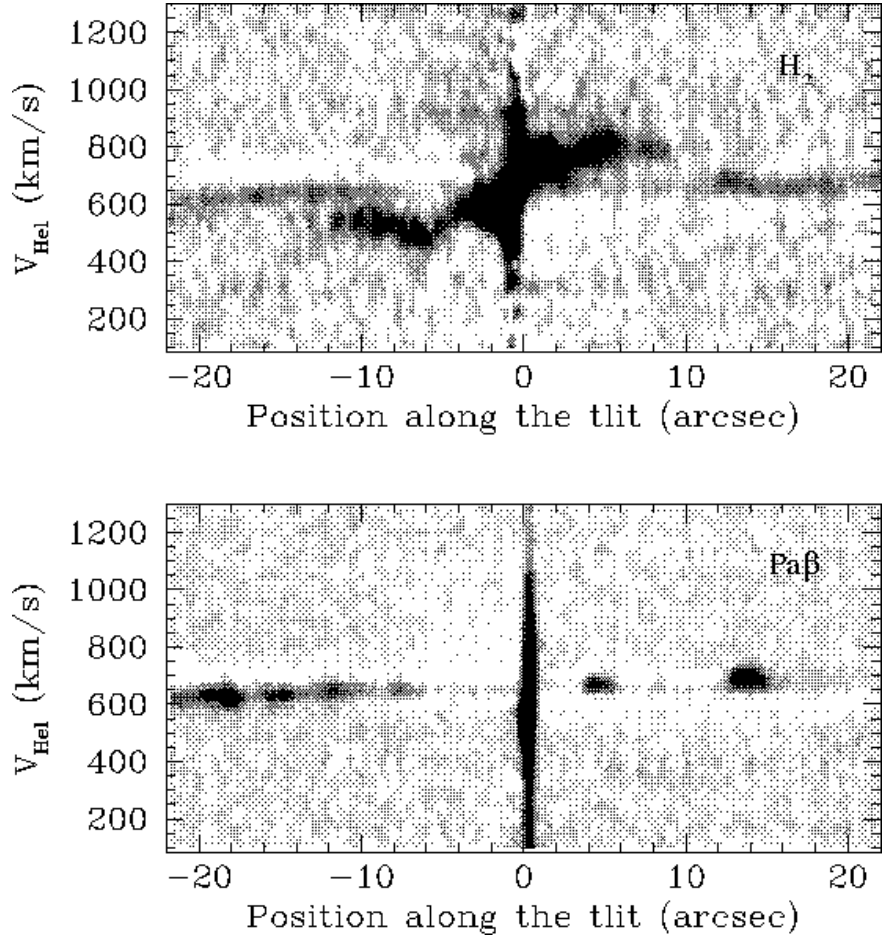,angle=0,width=0.7\linewidth} 
\end{center}
\caption{\label{fig:greyvel} Grey scales of \HMOL\ and \PB\ position-velocity
diagrams at NUC-2.}
\end{figure*}

\begin{figure*}
\begin{center}
\epsfig{figure=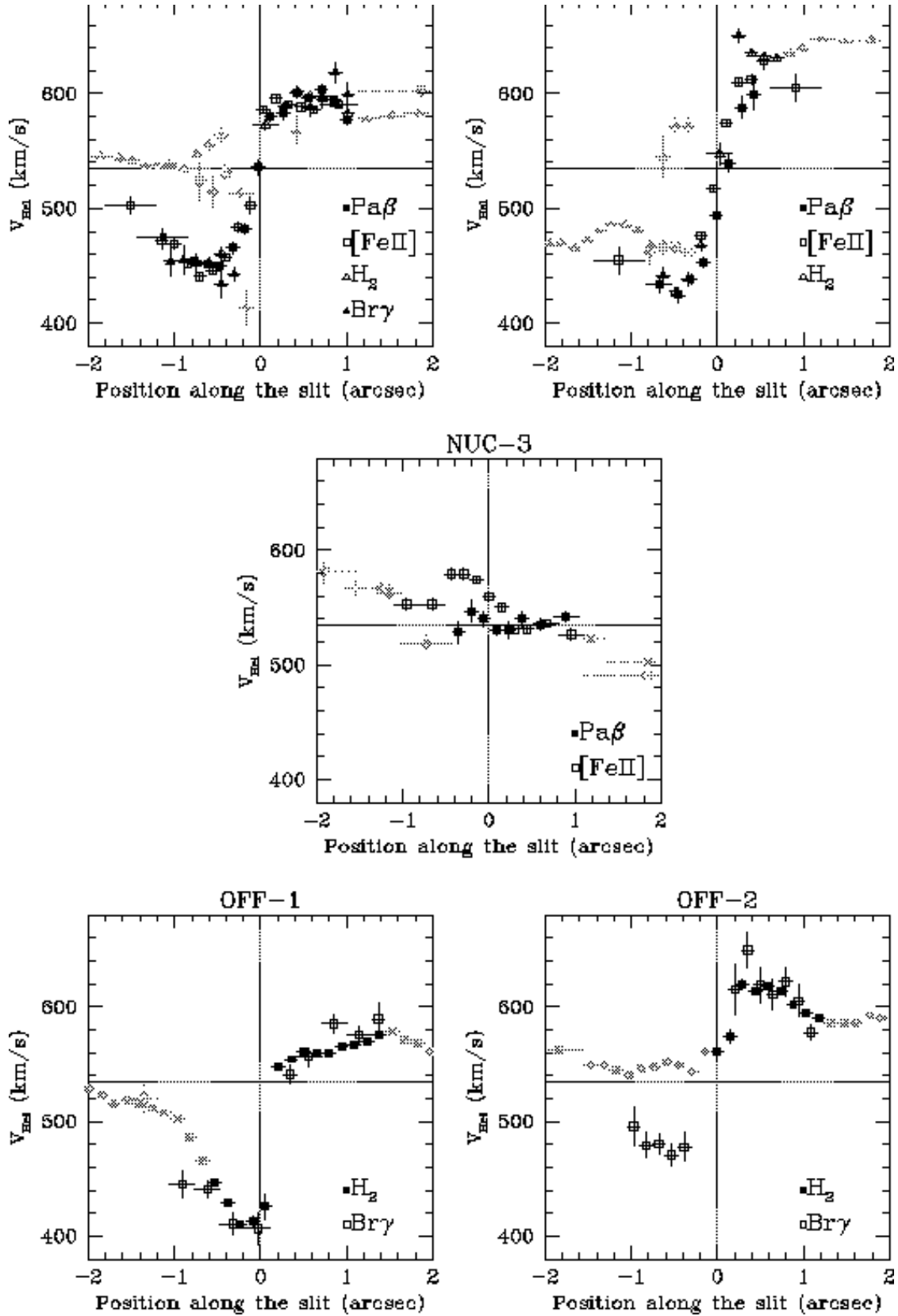,angle=0,width=0.9\linewidth} 
\end{center}
\caption{\label{fig:compare} Comparison 
of velocities in the inner 2\arcsec. The grey color marks the points 
"contaminated" by the extended component which are not used in the fit.}
\end{figure*}

\begin{figure*}
\centerline{
\epsfig{figure=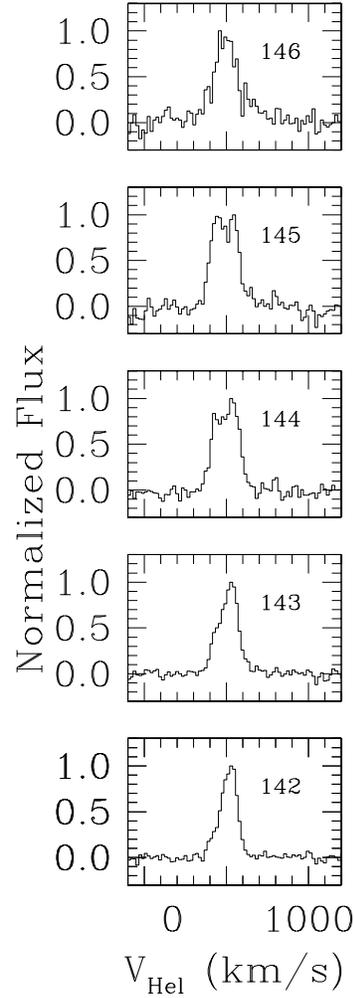,angle=0,width=0.7\linewidth}}
\caption{\label{fig:h2double} \HMOL\ line profiles along the slit at NUC-1
(the number in the upper right corner of each panel is the detector
coordinate at which the profile was extracted) showing the presence
of two different components. The blue one is the "nuclear" component.}
\end{figure*}

\begin{figure*}
\begin{center}
\begin{tabular}{c}
\epsfig{figure=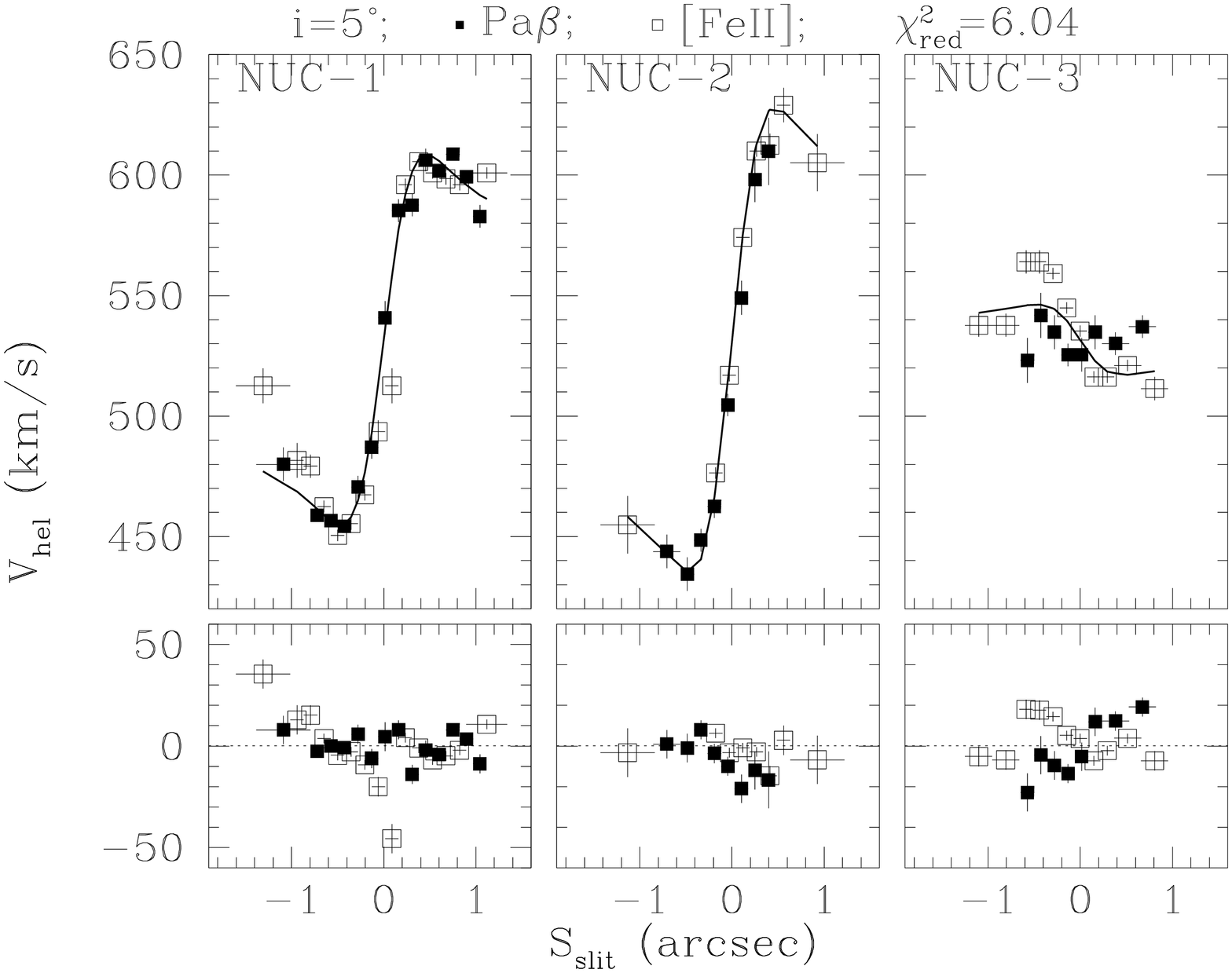,angle=-90,width=0.7\linewidth} \\
\\
\epsfig{figure=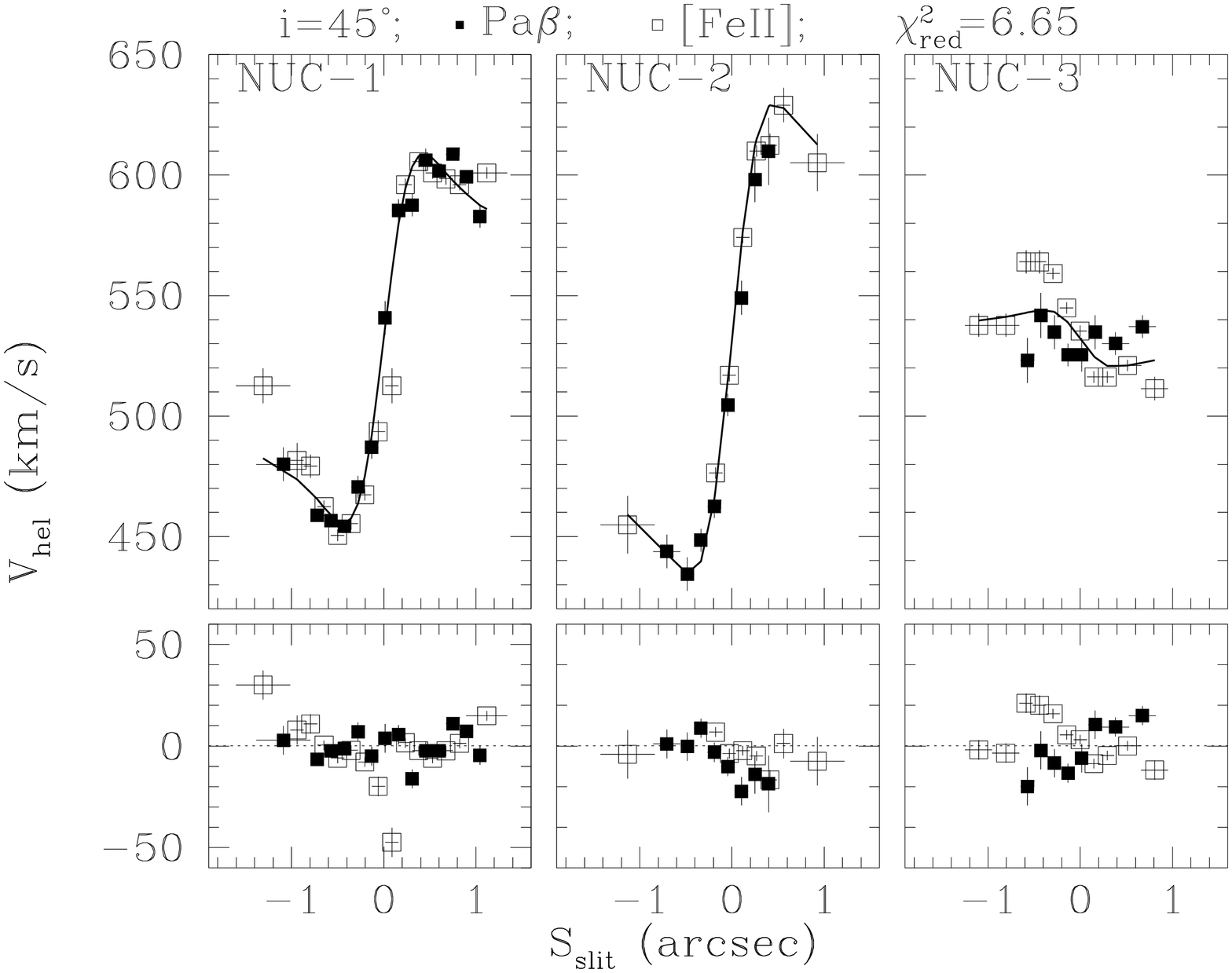,angle=-90,width=0.7\linewidth} \\
\end{tabular}
\end{center}
\caption{\label{fig:fit} Observed nuclear rotation curves (solid squares
are for \PB, empty squares for \FeII) with model fit (solid line). 
The lower insets in each panel are the residuals. The label at the top
of each panel indicates the assumed inclination and reduced \Chi\ value.}
\end{figure*}

\begin{figure*}
\begin{center}
\begin{tabular}{c}
\epsfig{figure=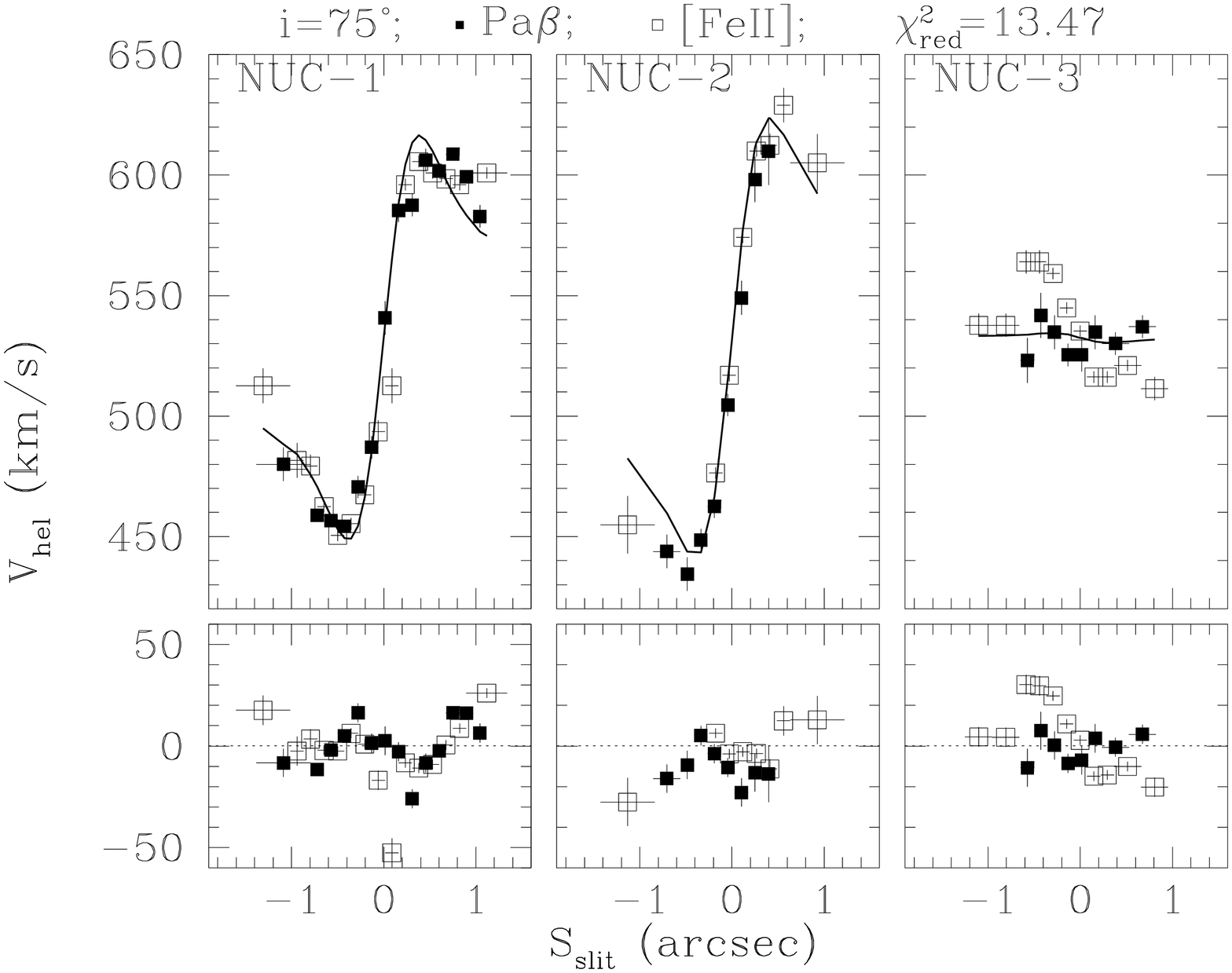,angle=-90,width=0.7\linewidth} \\
\\
\epsfig{figure=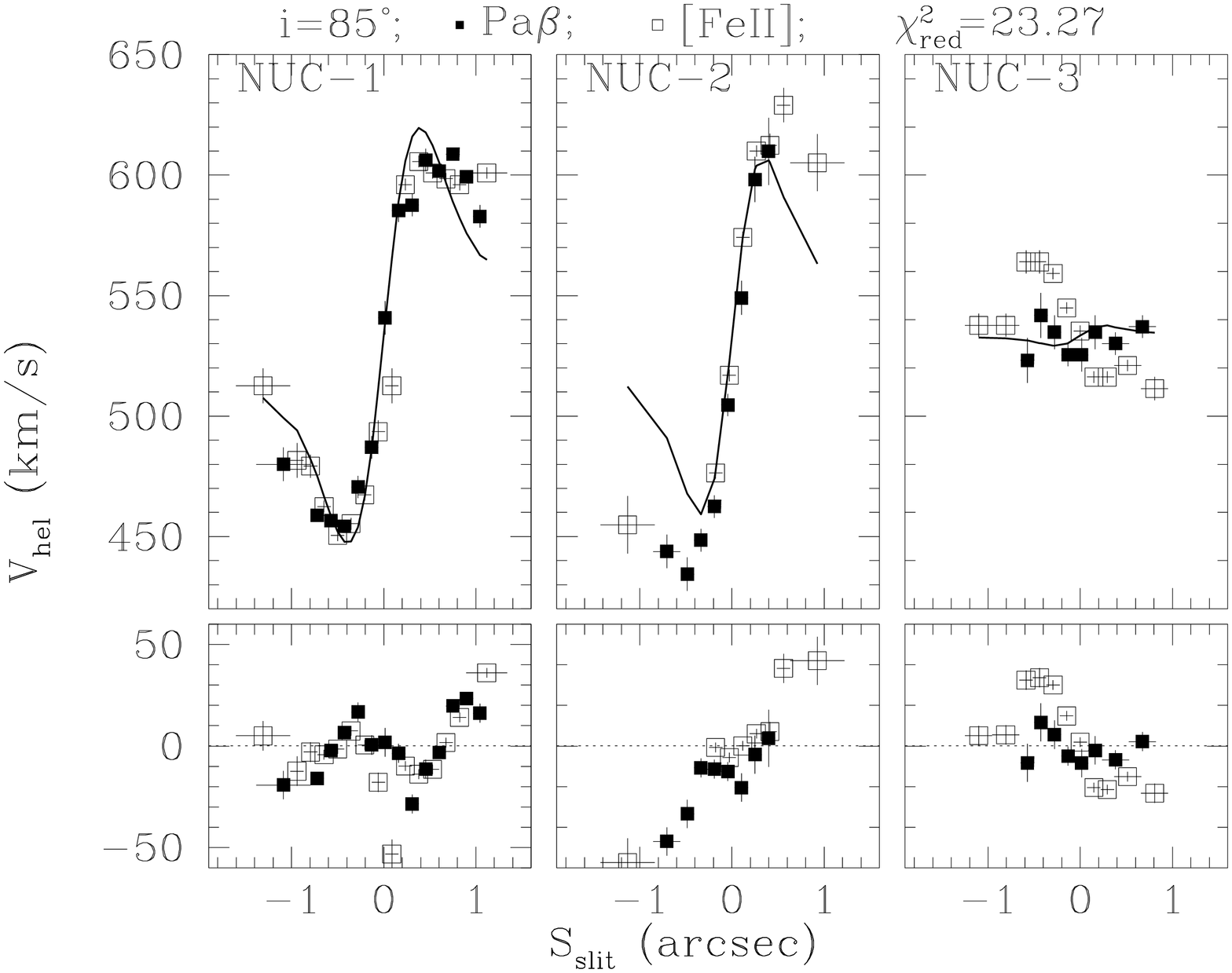,angle=-90,width=0.7\linewidth} \\
\end{tabular}
\end{center}
\caption{\label{fig:fit2} As in Fig. 12 but for values of the disk
inclination which do not give acceptable fits.}
\end{figure*}

\begin{figure*}
\begin{center}
\epsfig{figure=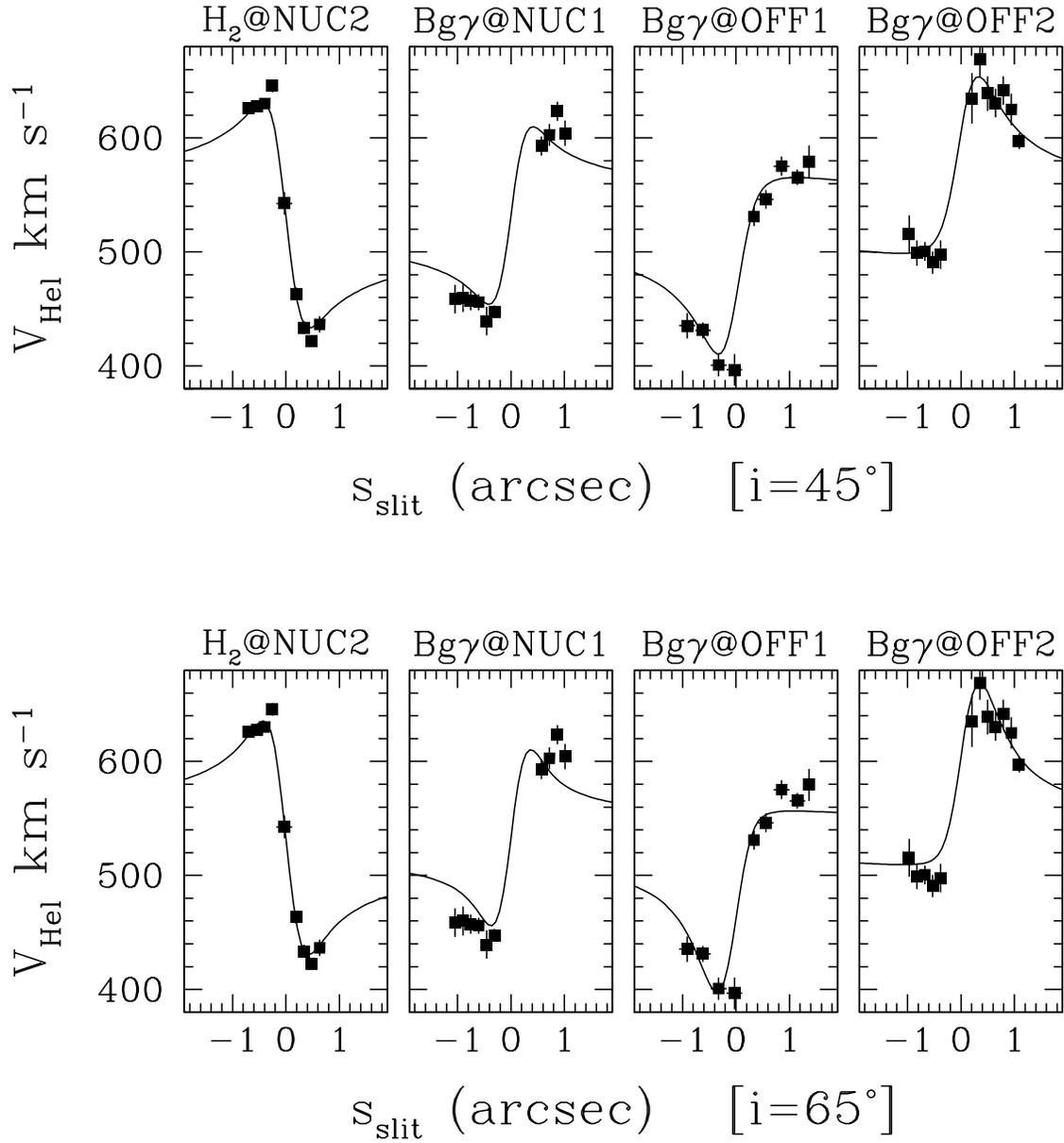,angle=0,width=0.9\linewidth} \
\end{center}
\caption{\label{fig:fith2bg} Model rotation curves compared with 
\BG\ and \HMOL\ data which were not used to constrain the fit.}
\end{figure*}

\begin{figure*}
\begin{center}
\epsfig{figure=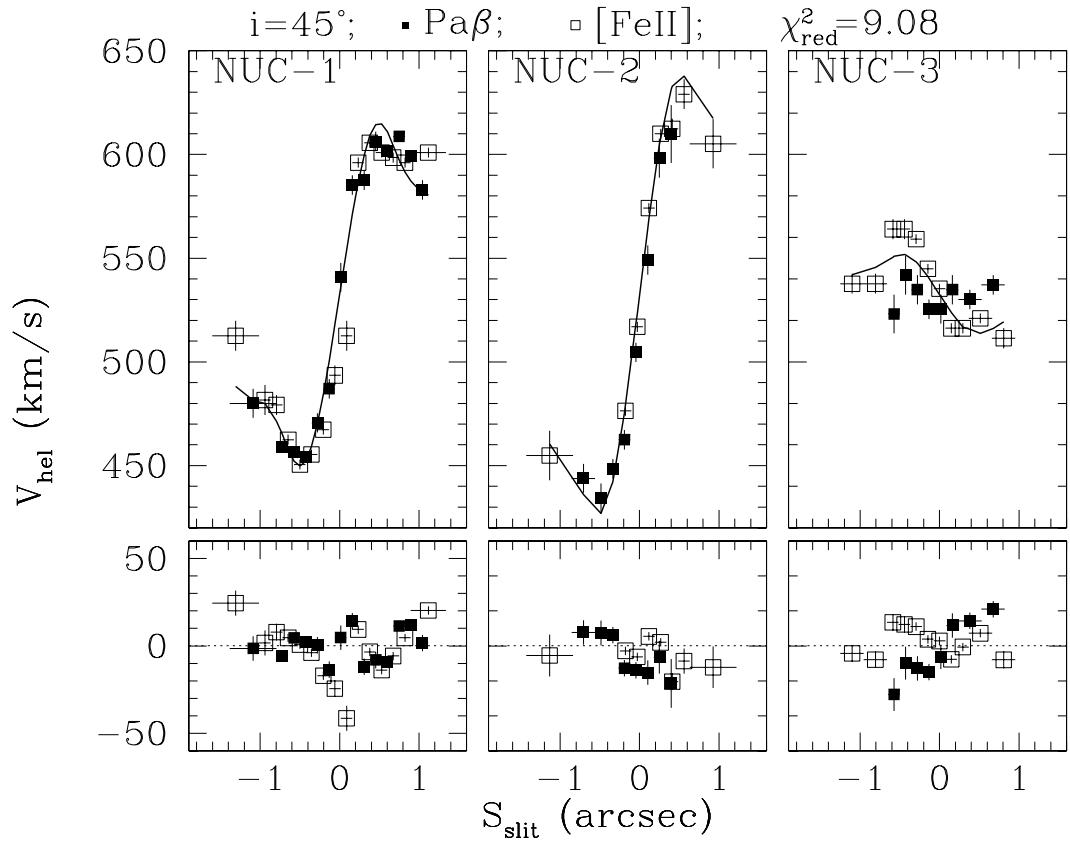,angle=-90,width=0.9\linewidth} 
\end{center}
\caption{\label{fig:fitint} Fit with the double exponential surface
brightness distribution for $\I=45\arcdeg$. Symbols as in Fig. 12.}
\end{figure*}

\begin{figure*}
\begin{center}
\begin{tabular}{c}
\epsfig{figure=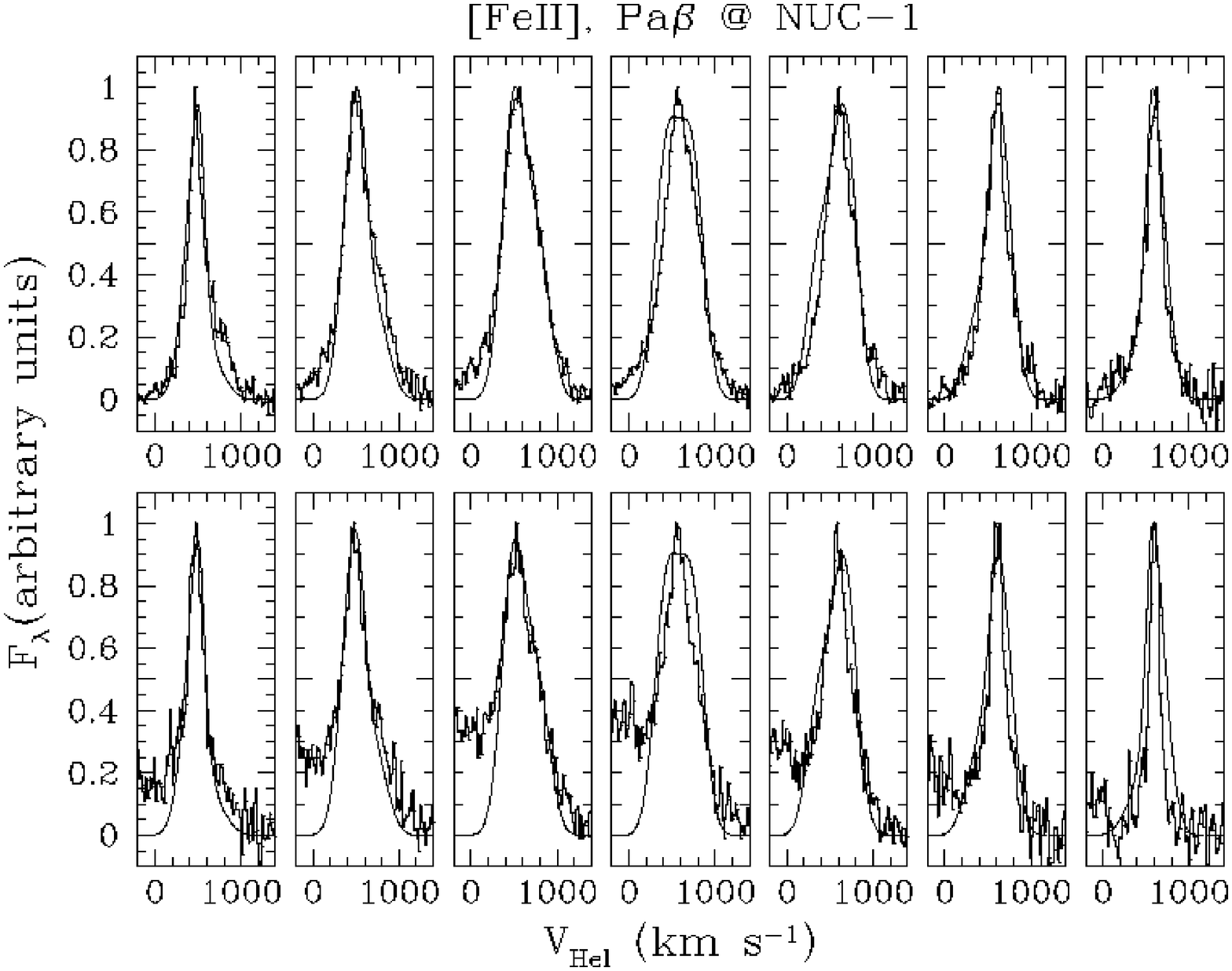,angle=-90,width=0.7\linewidth} \\
\epsfig{figure=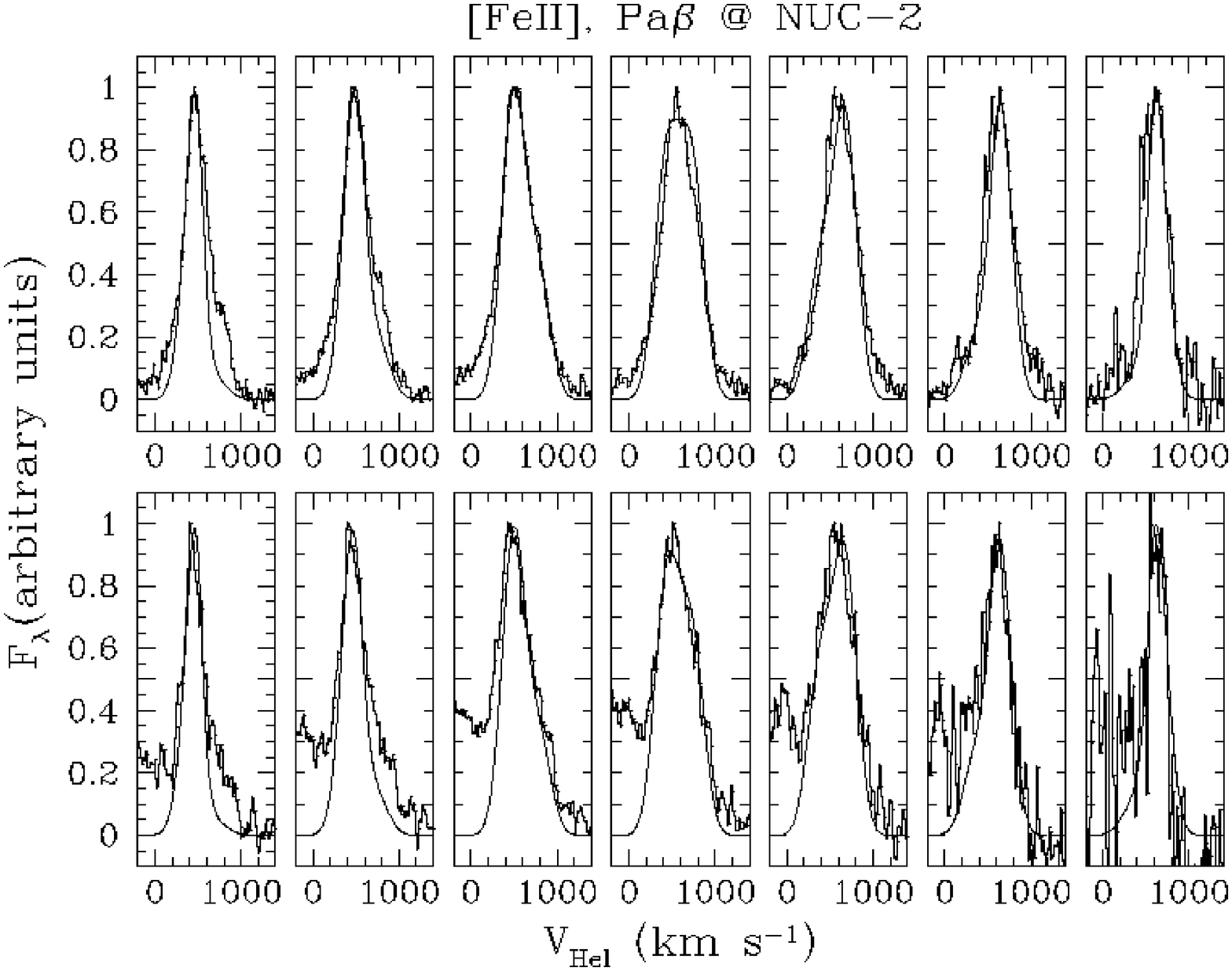,angle=-90,width=0.7\linewidth} \\
\end{tabular}
\end{center}
\caption{\label{fig:prof} Expected line profiles obtained
from a disk in keplerian rotation around a point mass compared
with the observed ones. The inclination of the disk is
\I=45\arcdeg, the surface brightness line
distribution is a double exponential (see Table 2 for the values of the
model parameters).}
\end{figure*}

We have showed that while the \HMOL\ rotation curve well fits our picture of a
fast rotating disk around a SMBH within the central 2\arcsec, at larger radii
the \HMOL\ velocities at NUC-2 are very different from those of the ionized
gas. From the inspection of Fig. \ref{fig:greyvel}, the large scale velocity
structure of the \HMOL\ shows two distinct velocity components.  One has a
constant velocity gradient and corresponds to the extended rotating structure
also seen in the ionized gas.  The second velocity component, only seen in
\HMOL\ between 2\arcsec\ and 10\arcsec, is reminiscent of a rotating ring with
an inner radius of 6\arcsec\ and a keplerian fall-off. Fig. \ref{fig:h2rot}
shows that the \HMOL\ flux at NUC-1 within 2\arcsec\ has two secondary maxima
at $\simeq 1\arcsec$ on both sides from the nucleus, as expected from a
circumnuclear ring.  The same two maxima, however, are not observed at NUC-2.
Comparing the projected radius at NUC-1 and the kinematical inner radius at
NUC-2, the "ring" must be almost edge-on ($\I>80\arcdeg$), in agreement
with the absence of the high velocity \HMOL\ components at NUC-1,
OFF-1 and OFF-2 where the slit PA is perpendicular.

In summary, this putative molecular "ring" has an inner radius of 
$\simeq 100\PC$, is almost edge on ($\I>80\arcdeg$) and,
from the velocity amplitude, includes a mass $M=0.6-2.5\xten{9}\Mo$.
The uncertainty on the mass derives from the uncertainties 
on the inclination and the inner radius of ring.

This high velocity component is the \HMOL\ counterpart of the 100\PC-scale
molecular ring previously identified by several authors in CO and FIR
\citep{israel90,israel91,ryd93,hawa93}.  In CO,
its major axis is oriented perpendicularly to the jet at PA$\sim 145\arcdeg$,
its radius ranges between 80 and 140\PC\ and its rotational velocity ($\simeq
220\KM\S\1$ at $\sim 120\PC$, rescaling from  3\MPC\ to 3.5\MPC\ distance)
results in an apparent dynamical enclosed mass of 1.5\xten{9}\Mo\
\citep{ryd93}.

A problem with the ring picture is that the \HMOL\ velocity decrease
observed between 6\arcsec\ and 10\arcsec\ cannot be ascribed to circular
keplerian motions.  The luminosity density plotted in Fig.  \ref{fig:nuker}b
and a Mass-to-Light ratio $\Upsilon\sim 1\Mo/\LKo$ naturally explain the mass
enclosed by the \HMOL\ ring within $R=6\arcsec$. However, the mass distribution
plotted in Figure \ref{fig:nuker}c is inconsistent with the observed velocity
decrease between 6\arcsec\ and 10\arcsec\ from the nucleus, which would require
negligible mass distributed in that range. Furthermore this model does
not explain the presence of the other extended component.

The observed velocity field could be readily understood in terms of the gas
dynamics of the gaseous bar observed by \citet{mira99}.  We note that the two
\HMOL\ velocity components form an approximate "figure-of-8" shape in the
position-velocity diagram of Fig.  \ref{fig:greyvel}. Similar velocity
structures have been observed in edge-on spiral galaxies believed to contain
barred potentials \citep{merr99,bureau}. Hydrodynamical and N-body
simulations \citep{ab00,ba00} suggest that these characteristic velocity fields
originate in weak barred potentials which support both the $x_1$ and $x_2$
families of orbits.  The slowly rotating component, seen in both ionized and
molecular gas, is likely produced by the gas in the disk outside the bar, in
combination with gas streaming along the $x_1$ family aligned with the axis of
the bar. The "anomalous" \HMOL\ component is generated by gas within the Inner
Lindblad Resonance (ILR) corresponding to the $x_2$ family of orbits. In most
of the models, a gas pseudo-ring forms at or close to the location of the outer
ILR (i.e. in our case the ring structure seen in \HMOL\ and CO). These models
however do not include a BH. In the presence of a weakly barred potential and a
supermassive BH, \citet{wada} and \citet{shlosh}  have shown that an additional
nuclear Lindblad resonance (NLR) is formed close to the nucleus. Numerical
results show that, on a dynamical time-scale, the gas is accumulated into a
nuclear ring within the NLR \citep{fu98}.  The gas ring becomes gravitationally
unstable when the surface density of the gas reaches a critical value,
fragments and forms a nuclear gas disk of several 10\PC-radius \citep{fu00}.
This could provide a natural explanation for the rapidly rotating nuclear disk
we see within the central 2\arcsec\ of Cen A, possibly accounting for the
discrepancy between the apparent line of nodes observed in \PA\ and the
kinematical one (end of Sec. \ref{sec:fit}).

A problem for the bar picture could arise from the fact that the bar is not
seen in the near infrared isophotes. As shown above, the "ring" system is
close to edge-on and, depending on its actual shape, the barred potential
might not be readily identified from the isophotes. In this case, it has been
suggested that boxy or peanut-shaped bulges in edge-on systems can be
associated to bars \citep[e.g.][]{merr99,ba00}.  However, even in the
near-IR, the analysis of Cen A isophotes is hindered by the high 
reddening and no
definite conclusion can be drawn.  Another possibility is that
the bar is a transient phenomenon, visible only in gas, as in the case of the
Circinus galaxy \citep{maio00}. This fact was suggested as a way to fuel
an active galactic nucleus \citep{shlosh}. 

Though the explanation involving a gaseous bar is very appealing, nonetheless
we cannot exclude that the \HMOL\ kinematics could be explained with a warped
thin disk as in the models by NBT92 and Q92: in this case the velocity decrease
of the ring-like \HMOL\ rotation curve at NUC-2 might be ascribed to the
twisting of the disk line of nodes and/or changing inclination.  The \HMOL\
ring thus represents the inner edge of the dust lane and the inner radius of
the ring might be identified with the tidal disruption radius of a molecular clouds in
the nuclear gravitational potential.  The radius for tidal disruption of a
spherical cloud in circular rotation around a spherical mass M is simply 
\begin{equation}
R_\mathrm{tidal}=\left(3\frac{M}{m_\mathrm{c}}\right)^\frac{1}{3}r_\mathrm{c} =
\left(\frac{9M}{4\pi\rho_\mathrm{c}}\right)^\frac{1}{3} = 134\PC
\left(\frac{M_9}{n_4}\right)^\frac{1}{3}
\end{equation}
where $m_\mathrm{c}$, $r_\mathrm{c}$ and $\rho_\mathrm{c}$ are the mass, radius
and density of the cloud, $M_9$ is the mass in units of \ten{9}\Mo\ and $n_4$
is the particle number density in units of \ten{4}\CM\3. Since the mass
enclosed within the ring is around $M_9\simeq 1$, the tidal radius of a
$\simeq\ten{4}\CM\3$ cloud is in excellent agreement with the observed inner
radius of the ring. This molecular gas density, apart for being a reasonable
value for a molecular cloud, is also similar to the value inferred by
\citet{israel90} from modeling the \HMOL\ emission.

In summary, even if the gaseous bar picture is very appealing in the framework
of the fueling of the AGN, the warped disk explanation cannot be excluded
and more data are required to prove the existence of a bar.

\section{\label{sec:summary}Summary}

New high spatial resolution, near-IR spectroscopy at several position angles
and locations within the central 20\arcsec\ of the nearest radio galaxy
Centaurus A have allowed us to derive velocities of ionized and molecular gas
(\FeII, \PB, \BG\ and \HMOL) that are not compromised by excitation effects or
obscuration.  We have identified three distinct kinematical systems: (i) a
"nuclear" rotating disk of ionized gas, confined within the inner 2\arcsec,
which is the counterpart of the \PA\ feature previously revealed by NICMOS
imaging on the HST; (ii) a "ring-like" system with a $\sim 6\arcsec$ inner
radius detected only in \HMOL\, the counterpart of the 100\PC\ scale "torus"
detected in CO by other authors; (iii) a normal, "extended" component of gas
rotating in the galactic potential.

The nuclear "disk", characterized by a low excitation spectrum, has an electron
density in the range 3.5-4.5\CM\3\ (in log) and its mass in ionized gas is
$\sim \ten{3}\Mo$. We have detected emission in the high ionization line
\SIX\WL 12523.5\AA; this is the first detection of a coronal line from an
object characterized by a low excitation spectrum.

The nuclear rotation curves are well modeled by a thin disk in keplerian
rotation around a central mass concentration. The line widths are consistent
with unresolved rotation of the disk and there is no evidence for strong
non-gravitational motions. The rotation curves suggest a central mass
concentration that is dark ($M/L\simeq 20\,\Mo/\LKo$) and point-like at the
spatial resolution of the data ($R<0\farcs25\sim 4\PC$) which we interpret as a
supermassive black hole. The minimum dynamical mass is $\M>6\xten{7}\Mo$.  If
we use the radio jet observations to exclude the case of a face-on nuclear disk
($\I<15\arcdeg$), we can constrain the mass to the range
$\M=2^{+3.0}_{-1.4}\xten{8}\Mo$.

The active nucleus in Centaurus A is accreting well below the Eddington limit,
as found in other FR-I radio sources. A comparison between the X-ray luminosity
and the FIR luminosity suggests that the AGN continuum lacks the big blue bump,
a feature which, together with the low accretion efficiency, suggests an
advection dominated accretion flow (ADAF).

The ratio of the BH mass to the galaxy mass is a factor of ten below the
average of values directly measured in other galaxies suggesting either that it
is overestimated or that it is characterized by a large spread. However, the BH
mass over galaxy mass ratio of Centaurus A agrees with predictions derived from
integrated quasar light and the same value measured for a sample of Seyfert 1
galaxies from reverberation mapping.

The ring-like system appears characterized by non-circular motions
and a "figure-of-8" pattern observed in the \HMOL\ position-velocity diagram 
might provide kinematical evidence for the presence of a nuclear bar.
However, with the present data, we cannot exclude that the same velocity
systems could be explained with a warped-disk geometry.

Centaurus A is the nearest example of an active galaxy whose nucleus is 
heavily obscured by gas and dust.  Thus, despite its relative proximity, it 
has not been possible to use high resolution optical spectroscopy to study 
the kinematics of the gas near the putative black hole at the center of Cen 
A. The advent of high sensitivity, high spatial resolution near-IR 
spectroscopy on 8m-class telescopes has made it possible to resolve the 
multiple velocity components of the gas in the vicinity of the nucleus, and 
to obtain a dynamical BH mass estimate for this obscured nucleus.  Since 
dust obscuration is relatively common, and is likely to be even more so at 
higher redshifts, even higher spatial resolution near-IR spectroscopy, 
either from space or via adaptive optics on the ground, will be essential 
to extend this work to other obscured AGNs.

\acknowledgments
We thank the User Support Group at ESO for performing the observations
and delivering the data.
We thank Roberto Maiolino, Tino Oliva, Leslie Hunt and Andrew Baker
for useful discussion.
A.M. acknowledge the partial support of the Italian Space Agency (ASI)
through the grant ARS--99--44 and of the Italian Ministry for 
University and Research (MURST) under grant Cofin98-02-32.
A.M., A.C. and DJA  acknowledge support from the STScI Visitor Program.

\appendix
\section{Appendix: Analytical formula for the rotation curve along the slit}

At a given slit position, the rotation curve can be written as a simple
analytical expression assuming negligible slit width and a thin
gaseous disk circularly rotating in a spherical gravitational potential $\Phi$.
The derivation of the formula is described in detail in Sec. 7 of
\citet{m87bh}.  The geometry and definitions are summarized in their Fig. 11
which we reproduce in Fig. \ref{fig:diskgeo}.

Briefly, we choose a reference frame $XY$ on the plane of the sky,
such that $X$ is aligned along the line of nodes of the disk
which has an inclination $i$ with respect to the
line of sight ($i=0$ is the face-on case); the slit
forms an angle $\theta$ with the line of nodes and
is at distance $b$ from the center of the disk (projected on the sky).
If $s$ is the coordinate along the slit and its zero point O is defined
as the point closer to the disk center, the $X$ and $Y$ coordinates
on the plane of the sky of a point on the slit are given by:
\begin{eqnarray}
\label{eq:kepdisk}
X & = & -b \sin\theta + s \cos\theta \nonumber 	\\
Y & = & ~b \cos\theta + s \sin\theta
\end{eqnarray}
and the deprojected radial distance from the disk center $R$ is
\begin{equation}
\label{eq:radius}
R^2 = X^2+\frac{Y^2}{(\cos i)^2}
\end{equation}
With a spherical gravitational potential $\Phi(R)$ the velocity for circular
orbits at distance $R$ from the nucleus is
\begin{equation}
V_\circ (R) = \left| R\frac{d\Phi}{dR} \right|^\frac{1}{2} =
\left( \frac{GM(R)}{R} \right)^\frac{1}{2}
\end{equation}
where $M(R)$ is the mass enclosed at radius $R$, a constant value $M$
in case of a point mass.
Therefore the expression for the velocity at a given slit position $s$ is
\begin{equation}
\label{eq:kepler}
V = V_{sys} - V_\circ (R)\sin i\frac{X}{R} =
V_{sys} - (GM)^{0.5}\sin i\frac{X}{R^{1.5}}
\end{equation}
This simple formula can also be used to obtain the expected
velocity along the slit when taking into account the spatial resolution 
of the observations and the finite slit width.
Choosing the reference frame described by $s$ and $b$ i.e.
the coordinate along the
slit and the impact parameter, we obtain
the model rotation curve $V_{psf}$:
\begin{equation}
V_{psf}(S) =
\frac{\int_{S-\Delta S}^{S+\Delta S} ds \int_{B-h}^{B+h} db
\int\int_{-\infty}^{+\infty} db^\prime ds^\prime
V(s^\prime,b^\prime) I(s^\prime,b^\prime) P(s-s^\prime, b-b^\prime) }
{ \int_{S-\Delta S}^{S+\Delta S} ds \int_{B-h}^{B+h} db
\int\int_{-\infty}^{+\infty} db^\prime ds^\prime
                    I(s^\prime,b^\prime) P(s-s^\prime, b-b^\prime) }
\end{equation}
where $V(s^\prime,b^\prime)$ is the keplerian velocity from
eq. \ref{eq:kepler},
$I(s^\prime,b^\prime)$ is the intrinsic luminosity distribution of the
line, $P(s^\prime-s, b^\prime-b)$ is the spatial PSF
along the slit direction.
$B$ is the impact parameter (measured at the center of the slit)
and $2h$ is the slit size, S is the position along the slit at which
the velocity is computed and $2\Delta S$ is the pixel
size.
For the PSF we can assume a gaussian with FWHM equal to the seeing of
observations $\sigma$ i.e.
\begin{equation}
P(s-s^\prime, b-b^\prime) = \frac{1}{\sqrt{2\pi\sigma^2}} \exp \left(
-\frac{1}{2}\frac{(s-s^\prime)^2}{\sigma^2}
-\frac{1}{2}\frac{(b-b^\prime)^2}{\sigma^2} \right)
\end{equation}

Note that the rotation curve obtained with a disk inclination \I\ and an
angle between the slit and the line of nodes \Th, can also be obtained with
angles $180\arcdeg-\I$ and $180\arcdeg+\Th$.

A similar formula gives the line profiles:
\begin{equation}
\Phi(v; S) =
\frac{\int_{S-\Delta S}^{S+\Delta S} ds \int_{B-h}^{B+h} db
\int\int_{-\infty}^{+\infty} db^\prime ds^\prime
\phi[v-V(s^\prime,b^\prime)] I(s^\prime,b^\prime) P(s-s^\prime, b-b^\prime) 
}
{ \int_{S-\Delta S}^{S+\Delta S} ds \int_{B-h}^{B+h} db
\int\int_{-\infty}^{+\infty} db^\prime ds^\prime
                    I(s^\prime,b^\prime) P(s-s^\prime, b-b^\prime) }
\end{equation}
where $\phi(v)$ is the convolution of the intrinsic line profile with
the instrumental response.

\clearpage

\begin{figure*}
\begin{center}
\begin{tabular}{c}
\epsfig{figure=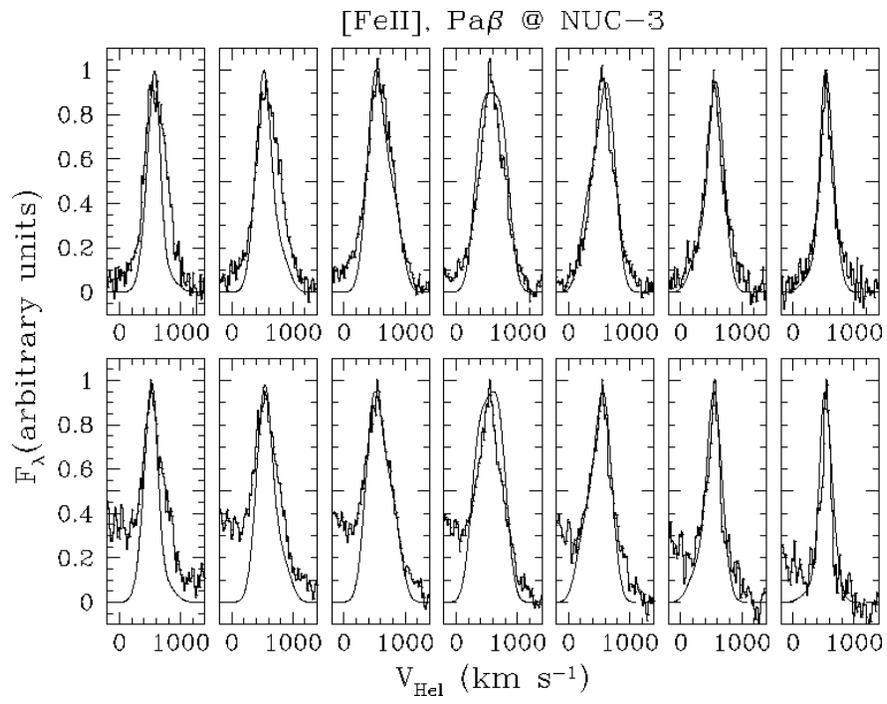,angle=-90,width=0.7\linewidth} \\
\end{tabular}
\end{center}
\caption{As in previous figure.}
\end{figure*}

\begin{figure*}
\begin{center}
\epsfig{figure=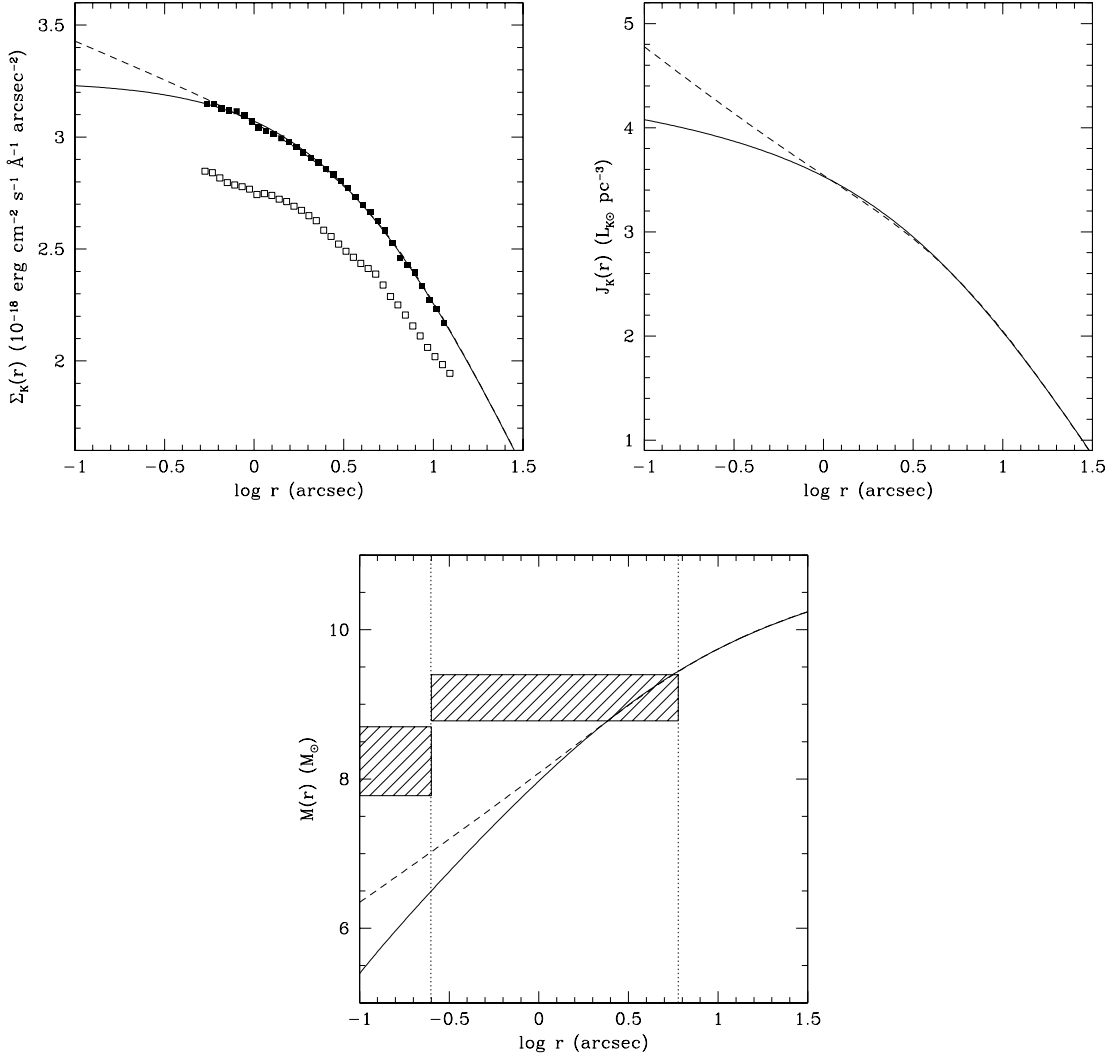,angle=0,width=0.9\linewidth}
\end{center}
\caption{\label{fig:nuker} From left to right: a) nuker-law fit of the observed 
K-band surface brightness profile from Paper II (solid line). The dashed line
is a fit with the steepest nuclear slope consistent with the data. The empty squares are the observed 
K band points. The filled squares are the same points after reddening
correction.
b) corresponding deprojected luminosity densities and c) masses
enclosed within radius $r$ with $\Upsilon=1\Mo/\LKo$. 
The dashed area on the left represents
the observed mass within 0\farcs25 radius and the other dashed area
is the enclosed mass within 6\arcsec\ required to explain the
\HMOL\ rotation curve of the ring-like feature at NUC-2
with a keplerian rotation.}
\end{figure*}

\clearpage
\begin{figure*}
\begin{center}
\epsfig{file=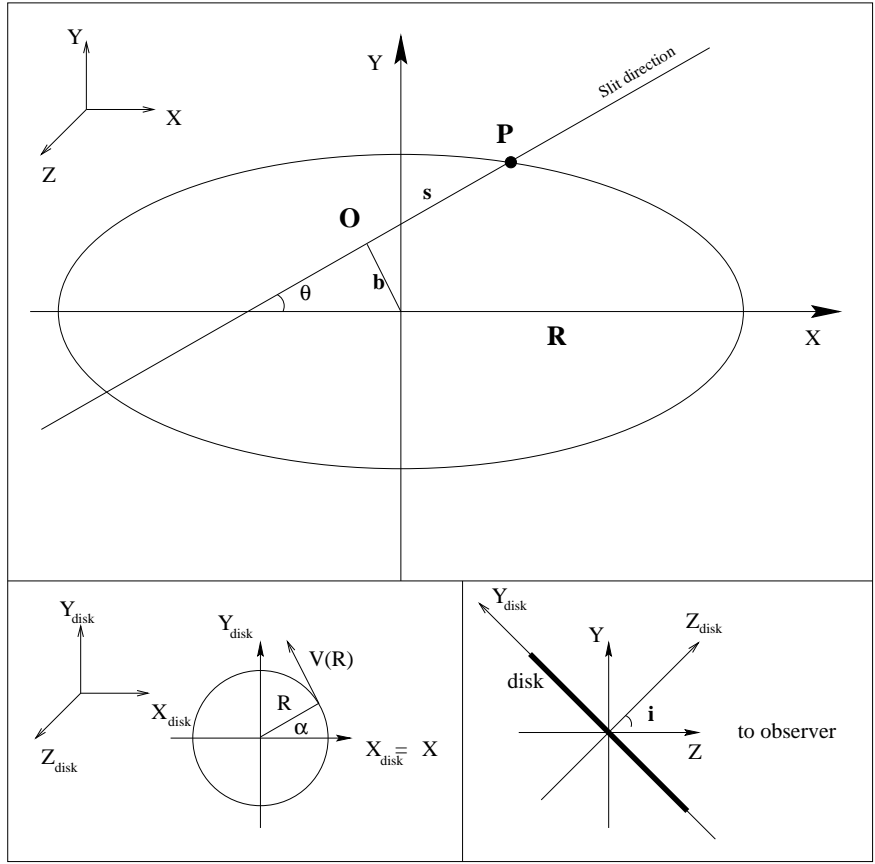,angle=0,width=0.7\linewidth}
\end{center}
\caption{\label{fig:diskgeo}
Schematic representation of the reference frames
used in the determination of the keplerian rotation curve. $XY$ is the
plane of the sky, $X$ is directed along the major axis of the disk,
and $Z$ is directed toward the observer. $X_{disk}Y_{disk}$ is the reference
frame on the disk plane such that $X_{disk}=X$
(Figure from Macchetto et al. 1997).}
\end{figure*}

\clearpage

\begin{deluxetable}{lccccr}
\tablewidth{0pt}
\tablecaption{\label{tab:log} Log of observations.}
\tablehead{
\colhead{Band} & \colhead{Lines} & \colhead{Target Name} & \colhead{PA (deg)} &
\colhead{Date (1999)} & \colhead{Seeing (\arcsec)} }
\footnotesize
\startdata
K & \HMOL, \BG	& NUCLEUS-1 	& 32.5	 	& May 30	&  0\farcs5 \\ 
K & \HMOL, \BG	& OFF-1     	& 32.5	 	& May 30	&  0\farcs6 \\ 
K & \HMOL, \BG	& OFF-2     	& 32.5	 	& May 30	&  0\farcs7 \\ 
K & \HMOL, \BG	& NUCLEUS-2 	& 135.5  	& Jun  3	&  0\farcs4 \\ 
K & \HMOL, \BG	& OFF-2     	& 32.5	 	& Jun  3	&  0\farcs6 \\ 
J & \FeII, \PB	& NUCLEUS-1 	& 32.5	 	& Jun 23	&  0\farcs6 \\ 
J & \FeII, \PB	& NUCLEUS-1 	& 32.5	 	& Jul 22	&  0\farcs4 \\ 
J & \FeII, \PB	& NUCLEUS-2 	& -44.5  	& Jul 22	&  0\farcs4 \\ 
J & \FeII, \PB	& NUCLEUS-3 	& 83.5   	& Jul 22	&  0\farcs5 \\ 
\enddata
\end{deluxetable}

\begin{deluxetable}{cccccccccccc}
\tablewidth{0pt}
\tablecaption{\label{tab:fit} Fit Results\tablenotemark{a}.}
\tablehead{
\colhead{\I\tablenotemark{b}} &
\colhead{\M\tablenotemark{c}} &
\colhead{PA LoN\tablenotemark{d}} &
\colhead{\Vs\tablenotemark{e}} &
\colhead{\Chi\tablenotemark{f}} &
		& &
\colhead{\I\tablenotemark{b}} &
\colhead{\M\tablenotemark{c}} &
\colhead{PA LoN\tablenotemark{d}} &
\colhead{\Vs\tablenotemark{e}} &
\colhead{\Chi\tablenotemark{f}}
}
\startdata
\multicolumn{5}{c}{Constant line surface brightness} & & &
\multicolumn{5}{c}{Double Exponential line surface brightness}\\
  5 & 4357 & -14.1 & 532 & 6.0 & & &  5 &3832 & -15.0 & 532 &  8.4  \\
 15 &  507 & -14.3 & 532 & 6.0 & & & 15 & 470 & -15.4 & 532 &  9.2  \\
 25 &  202 & -14.0 & 532 & 6.1 & & & 25 & 177 & -16.0 & 532 &  8.7  \\
 35 &  122 & -13.6 & 532 & 6.3 & & & 35 & 102 & -16.1 & 532 &  8.7  \\
 45 &   93 & -13.2 & 532 & 6.7 & & & 45 &  72 & -15.5 & 532 &  9.1  \\
 55 &   89 & -11.8 & 533 & 7.5 & & & 55 &  59 & -13.5 & 533 & 10.0  \\
\hline
 65 &  105 & -10.6 & 533 & 9.3 & & & 65 &  55 & -14.0 & 534 & 12.5  \\
 75 &  181 & -7.8  & 533 &13.5 & & & 75 &  59 & -11.5 & 535 & 15.4  \\
 85 &  813 & -2.6  & 533 &23.3 & & & 85 &  68 & -17.2 & 535 & 24.4  \\
\enddata
\tablenotetext{a}{ Errors on parameters are $\Delta\M/\M\pm 20\%$,
$\Delta\Th\pm 5\arcdeg$, $\Delta\Vs\pm 5\KM\S\1$}
\tablenotetext{b}{ Degrees; $i=90\arcdeg$ edge-on; $i=0\arcdeg$ face-on.}
\tablenotetext{c}{ Units of \ten{6}\Mo.}
\tablenotetext{d}{ Degrees from North to East (LoN is the disk line of nodes).}
\tablenotetext{e}{ \KM\S\1}
\tablenotetext{f}{ Reduced $\chi^2$.}
\end{deluxetable}


\begin{thebibliography}{}
%
\bibitem[Alonso-Herrero et al.(1997)]{alonso}
Alonso-Herrero A., Rieke M.J., Rieke G.H., Ruiz M., 1997 \apj, 482, 747
%
\bibitem[Athanassoula \& Bureau(2000)]{ab00}
Athanassoula E., Bureau M., 2000, \apj, 522, 699
%
\bibitem[Bland, Taylor \& Atherton(1987)]{bta87}
Bland J., Taylor K., Atherton P.D., 1987, \mnras, 228, 595
%
\bibitem[Binette(1998)]{binette}
Binette L., 1998, \mnras, 294, L47
%
\bibitem[Bureau \& Freeman (1999)]{bureau}
Bureau M., Freeman K.C., 1999, \aj, 118, 126
%
\bibitem[Bureau \& Athanassoula(2000)]{ba00}
Bureau M., Athanassoula E., 2000, \apj, 522, 686
%
\bibitem[Capetti et al.(2000)]{cap00}
Capetti A., Schreier E.J., Axon D.J., Young S., Hough J.H., Clark S.,
Marconi A., Macchetto F.D., Packham C., 2000, \apj, in press
%
\bibitem[Chiaberge, Capetti \& Celotti(1999)]{chiab}
Chiaberge M., Capetti A., Celotti A., 1999, \aap, 349, 77
%
\bibitem[De Koff et al.(1996)]{dekoff96} 
De Koff S. , Baum S.A., Sparks W.B., Biretta J., Golombek D., 
Macchetto F.D., McCarthy P., and Miley G.K., 1996, ApJS 107, 621 
%
\bibitem[Devillard(1997)]{eclipse}
Devillard, N., 1997, The Messenger 87, 19 
%
\bibitem[Di Matteo et al.(2000)]{dimatteo}
Di Matteo T., Quataert E., Allen S.W., Narayan R., Fabian A.C.,
2000, \mnras 311, 507
%
\bibitem[Dufour et al.(1979)]{dufour}
Dufour R.J., van den Bergh S., Harvel C.A., et al, 1979 \aj, 84, 284
%
\bibitem[Eckart et al.(1990)]{eckart}
Eckart A., Cameron M., Rothermel H., et al, 1990, \apj, 363, 451
%
\bibitem[Elvis et al.(1994)]{elvis}
Elvis M., Wilkes B.J., McDowel J.C., et al, 1994, \apjs, 95, 1
%
\bibitem[Fabian \& Iwasawa(1999)]{fabian}
Fabian A.C., Iwasawa K., 1999, \mnras, 303, L34
%
\bibitem[Ford et al.(1997)]{ford97}
Ford H.C., Tsvetanov Z.I., Ferrarese L., Jaffe W., 1997, in IAU Symp. 184 
"The Central Regions of the Galaxy and Galaxies",  ed. Yoshiaki Sofue, 
(Dordrecht: Kluwer), p. 377
%
\bibitem[Fukuda, Wada \& Habe(1998)]{fu98}
Fukuda H., Wada K., Habe A., 1998, \mnras, 295, 463  
%
\bibitem[Fukuda, Habe \& Wada(2000)]{fu00}
Fukuda H., Habe A., Wada K., 2000, \apj, 529, 109
%
\bibitem[Hawarden et al.(1986)]{hawa93}
Hawarden T.G., Sandell G., Matthews H.E., et al, 1993, \mnras, 260, 844
%
\bibitem[Ho(1998)]{ho98}
Ho L.C., 1998, in "Observational Evidence for Black Holes in the Universe",
ed. S.K. Chakrabarti (Dordrecht: Kluwer), p. 157
%
\bibitem[Hui et al.(1993)]{hui}
Hui X., Ford H.C., Ciardullo R., Jacoby G.H., 1993 \apj, 414, 463
%
\bibitem[Kinzer et al.(1995)]{kinzer}
Kinzer R.L., Johnson W.N., Dermer C.D., et al, 1995, \apj, 449, 105
%
\bibitem[Kurpiewski \& Jaroszynski(2000)]{kup00}
Kurpiewski A., Jaroszynski M., 2000, ACTA Astr. in press (astro-ph/0003418)
%
\bibitem[Israel et al.(1990)]{israel90}
Israel F.P., van Dishoeck E.F., Baas F., et al, 1990, \aap, 227, 342
%
\bibitem[Israel et al.(1991)]{israel91}
Israel F.P., van Dishoeck E.F., Baas F., et al, 1991, \aap, 245, L13
%
\bibitem[Israel(1998)]{israel}
Israel F.P., 1998 \aapr, 8, 237
%
\bibitem[Kormendy \& Richstone(1995)]{kr95}
Kormendy J., Richstone D., 1995, \araa, 33, 581
%
\bibitem[Macchetto et al.(1997)]{m87bh}
Macchetto F.D., Marconi A., Axon D.J., Capetti A., Sparks W.B., Crane P., 
1997, \apj, 489, 579
%
\bibitem[Macchetto(1999)]{macchetto99}
Macchetto F.D., 1999, in "Towards a New Millennium in Galaxy
Morphology", D.L. Block et al. eds.
(Kluwer:Dordrecht), in press (astro-ph/9910089)
%
\bibitem[Magorrian et al.(1998)]{magorrian}
Magorrian J., Tremaine S., Richstone D., et al, 1998, \aj, 115, 2285
%
\bibitem[Maiolino et al.(2000)]{maio00}
Maiolino R., Alonso-Herrero A., Anders S., et al, 2000, \apj, 531, 219
%
\bibitem[Malkan, Gorjian \& Tam(1998)]{malkan}
Malkan M.A., Gorjian V., Tam R., 1998, \apjs, 117, 25
%
\bibitem[Maraston(1998)]{maraston}
Maraston C., 1998, \mnras, 300, 872
%
\bibitem[Marconi et al.(1996)]{clr1068}
Marconi A., van der Werf P.P., Moorwood A.F.M., Oliva E., 1996, \aap, 315, 335
%
\bibitem[Marconi et al.(2000)]{marconi00}
Marconi A., Schreier E.J., Koekemoer A., Capetti A., Axon D.J., Macchetto F.D.,
Caon N., 2000, \apj, 528, 276
%
\bibitem[Martel et al.(1999)]{martel99} 
Martel A.R., Baum S.A., Sparks W.B., Wyckoff E., Biretta J.A., 
Golombek D., Macchetto F.D., de Koff S., McCarthy P.J., Miley G.K., 
1999, ApJS 122, 81 
%
\bibitem[Mathieu \& Dejonghe(1999)]{mathieu}
Mathieu A., Dejonghe H., 1999, \mnras, 303, 455
%
\bibitem[Merrifield \& Kuijken(1999)]{merr99}
Merrifield M.R., Kuijken K., 1999, \aap, 345, L47
%
\bibitem[Merrifield, Forbes \& Terlevich(2000)]{merr00}
Merrifield M.R., Forbes D.A., Terlevich A.I., 2000 \mnras, 313, L29
%
\bibitem[Mirabel et al.(1999)]{mira99}
Mirabel I.F., Laurent O., Sanders D.B., et al, 1999 \aap, 341, 667
%
\bibitem[Mobasher et al.(1999)]{mobasher}
Mobasher B., Guzman R., Aragon-Salamanca A., Zepf S., 1999, \mnras, 304, 225
%
\bibitem[Moorwood \& Oliva(1988)]{moorwood}
Moorwood A.F.M., Oliva E., 1988 \aap, 203, 278
%
\bibitem[Moorwood et al.(1999)]{isaac}
Moorwood A.F.M., et al, 1999, The Messenger 95, 1
%
\bibitem[Moriondo, Giovanardi \& Hunt(1998)]{moriondo}
Moriondo G., Giovanardi C., Hunt L.K., 1998, \aap, 339, 409
%
\bibitem[Narayan, Kato \& Honma(1997)]{narayan}
Narayan R. Kato S., Honma F., 1997, \apj 476, 49
%
\bibitem[Nicholson, Bland-Hawthorn \& Taylor(1992)]{nbt92}
Nicholson R.A., Bland-Hawthorn J., Taylor K., 1992, \apj, 387, 503 (NBT92)
%
\bibitem[Oliva \& Origlia(1992)]{oliva92}
Oliva E., Origlia L. 1992, \aap, 254, 466
%
\bibitem[Oliva(1997)]{oliva97}
Oliva E., 1997, in IAU Colloquium 159 "Emission lines in active galaxies:
new methods and techniques" ASP Conf. Ser. 113, B.M. Peterson, F.Z. Cheng
A.S. Wilson eds., p. 288 
%
\bibitem[Oliva et al.(1994)]{clr}
Oliva E., Salvati M., Moorwood A.F.M., Marconi A., 1994, \aap, 288, 457
%
\bibitem[Pryor \& Meylan(1993)]{gc}
Pryor C., Meylan G., 1993, in ASP Conf. 150, "Structure and Dynamics
of Globular Clusters", eds. S.G. Djorgovski, G. Meylan (San Francisco ASP), 357 
%
\bibitem[Quillen et al.(1992)]{q92}
Quillen A.C., de Zeeuw P.T., Phinney E.S., Phillips T.G., 1992 \apj, 391, 
121 (Q92)
%
\bibitem[Quillen, Graham \& Frogel(1993)]{q93}
Quillen A.C., Graham J.R., Frogel J.A., 1993 \apj, 412, 550
%
\bibitem[Regan \& Mulchaey(1999)]{regan}
Regan M.W., Mulchaey J.S., 1999, \aj, 117, 267 
%
\bibitem[Richstone et al.(1998)]{rich98}
Richstone D., Ajhar E.A., Bender R., et al, 1998, \nat, 395, A14
%
\bibitem[Rydbeck et al.(1993)]{ryd93}
Rydbeck G., Wiklind T., Cameron M., et al, 1993, \aap, 270, L13
%
\bibitem[Schreier et al.(1996)]{schreier96}
Schreier E.J., Capetti A., Macchetto F., Sparks W.B., Ford H.J.,
1996 \apj, 459, 535
%
\bibitem[Schreier et al.(1998)]{schreier98}
Schreier E.J., Marconi A., Axon D.J., Caon N., Macchetto D., Capetti A.,
Hough J.H., Young S., Packham C., 1998, \apj, 499, L143 (Paper~I)
%
\bibitem[Shloshman, Frank \& Begelman(1989)]{shlosh}
Shloshman I., Frank J., Begelman M.C., 1989, \nat, 338, 45
%
\bibitem[Soria et al.(1996)]{soria96}
Soria R., Mould J.R., Watson A.M., et al, 1996, \apj, 465, 79
%
\bibitem[Tingay et al.(1998)]{tingay}
Tingay S.J., Jauncey D.L., Reynolds J.E., et al, 1998, \aj, 115, 960 
%
\bibitem[Tonry(1991)]{tonry91}
Tonry J.L., 1991, \apj 373, L1
%
\bibitem[Turner et al.(1997)]{turner97}
Turner T.J., George I.M., Mushotzky R.F., Nandra K., 1997, \apj, 475, 118
%
\bibitem[Unger, Clegg \& Stacey(2000)]{unger00}
Unger S.J., Clegg P.E., Stacey G.J., 2000, \aap, 355, 885
%
\bibitem[van der Marel et al.(1997)]{marel}
van der Marel R.P., de Zeeuw P.T., Rix H.W., Quinlan G.D., 
1997, \nat 385, 610
%
\bibitem[van der Marel(1997)]{marel97}
van der Marel R.P., 1997, in "Galaxy Interactions at Low and High Redshift",
IAU Symposium 186, D.B. Sanders, J. Barnes eds, Kluwer Academic Publishers, p. 333
%
\bibitem[van der Marel(1999)]{marel99}
van der Marel R.P., 1999 \aj, 117, 744
%
\bibitem[Wada(1994)]{wada}
Wada K., 1994, PASJ, 46, 165
%
\bibitem[Wandel(1999)]{wandel}
Wandel A., 1999, \apj, 519, L39
%
\bibitem[Wilkins \& Axon(1992)]{longslit92}
Wilkins T.W., Axon D.J., 1992, in ASP Conf. 25, Astronomical Data Analysis
Software and Systems I, eds. D.M. Worral, C. Biemesderfer, \& J. Barnes
(San Francisco: ASP), 427
\bibitem[Wilkinson et al.(1986)]{wilk86}
Wilkinson A., Sharples R.M, Fosbury R.A.E., Wallace P.T., 1986, \mnras 218, 297
%
\bibitem[Zhang \& Pradhan (1995)]{feiidata}
Zhang H.L., Pradhan A.K., 1995, \aap, 293, 953
\end{thebibliography}
\end{document}